\documentclass{aa}
\usepackage{epsfig,latexsym,longtable,lscape}

\begin{document}

\thesaurus{3
                 (11.19.2;  
                  11.19.5;  
                  11.19.7)}  

\title{Optical Surface Photometry of a Sample of Disk Galaxies. II Structural Components}

\author{M. Prieto$^{1}$, J. A. L. Aguerri$^{1,2}$,A. M. Varela$^{1}$ \& C. Mu\~noz-Tu\~n\'on$^{1}$}

\institute{1.- Instituto de Astrof\'\i sica de Canarias, E-38200 La Laguna, Tenerife, Spain \\
2.- Astronomisches Institut der Universitat Basel,  CH-4102 Binningen, Switzerland}   

\date{Received ; accepted }
  \authorrunning{M. Prieto et al}  

 \titlerunning{Structural components}
   \maketitle

\begin{abstract}
This work presents the structural 
decomposition of a sample of 11 disk galaxies, which span a range of different 
morphological types.      
 The $U$, $B$, $V$, $R$, and $I$ photometric 
information given in Paper I
 (color and color-index images and luminosity, 
ellipticity, and position-angle profiles) has been used to decide
 what types  of components  form the galaxies
before carrying out the decomposition.  We find and model such components as 
bulges, disks, bars, lenses and rings. 

\end{abstract}


\section{Introduction}

The fit of  structural models to the luminosity distribution of 
spiral galaxies is a complicated task. There are several difficulties,
among which we  emphasize the following: 
\begin{description}
\item 1) The presence of  components other the  bulge and disk. Sometimes these 
 structures are
clearly present in the form of, for example, large bars or lenses and we can model them. Often times
 this is not
possible particularly at short wavelengths  where the 
spiral arms, star formation
regions and the dust extinction can modify the  aspect of the smooth light distribution 
  due to the old population that  contributes most of the mass. 
\item 2) Bulge models are  not unique. All the
light-profile functions proposed for ellipticals can be used for the bulges of spirals.
 In recent years the generalized exponential bulge (Sersic 1968; Sparks 1988)
 has been the  most extensively used  model and this law is
 intrinsicly variable. 
\item 3) Although exponential disks are well established, it is necessary to have photometric data at large radii in order to avoid 
 contamination from other more central components (Prieto {\em et al.} 1992b). Most of the images
  of  nearby galaxies do not reach the low brightness levels to achieve this.
 \end{description}
  In general, the $B/D$ relation has very little reliability due to these above-mentioned
  reasons and also because the bulge scale length is sometimes smaller than the seeing disk. 
 Perhaps the most reliable decompositions are those for galaxies with a very well
 defined disk. This means, well  sampled  and free from other emission features.

In recent decades much work has been done in order to improve the methods for ascertaining the principal (bulge and disk)
components of disk galaxies from their optical photometric profiles. 
   Kormendy
(1977),  Boroson's (1981), Shombert \& Bothum (1987), Kent (1987), 
 Capaccioli, Held, \& Nieto (1987), Andredakis \& Sanders (1994),
De Jong (1996), Moriondo, Giovanardi, \& Hunt (1997).
  All these studies
  have resulted in gradual progress in this difficult task.
  
   The majority of 
  the above-mentioned studies   analyze  the bulge and disk without 
  considering the 
 presence of other components.
 In this article we try to provide a further  step in the unraveling of the
 complicated structures of  spiral galaxies, by taking into account all other
  components  which may be 
present in disk 
 galaxies. This is done by using the multi-color
 photometric information ( Aguerri {\em et al.} (1999), here after Paper I), to determine what 
  the different components are
 that comprise each galaxy before the decomposition  is attempted.
 We do this for a limited sample of 11 galaxies.
In paper I, 
we presented  the observations and data reduction
 for this sample of galaxies. 
 
\section {Model Fitting}

We use azimuthally averaged profiles to determine the various components.  This preserves information on non-axisymmetics features, such as bars $(m=2)$, because isophotal ellipse  fitting   allows 
  the
ellipticity and the position angle of the $m=2$ structures  to be fitted, 
as reflected in the
ellipticity and position angle profiles (Wozniak {\em et al.} 1995; Prieto  {\em et al.}
 1997; Varela, Mu\~noz-Tu\~n\'on, \& Simmoneau 1996). However when
the bar is very large and strong, we characterize the bulge and disk
 not by  azimuthally averaging but by individual 
profiles transverse  to the projected bar.

 The profiles are 
built from  ellipses of variable ellipticity and position
angle. To consider this variation is a way to correct for warps or any 
large-scale disk or bulge
deformation. 
The small variations in the luminosity of the disk are
thus averaged and the S/N increased considerably. 
  Points in the profile affected by additional structures, like rings or tight 
spiral arms, can be easily be omitted.

Moriondo {\em et al.} (1997) compared the 1D and 2D techniques and concluded that
the 1D
technique is always less accurate and sometimes inconsistent with 2D one and that  the inferior
quality of the 1-D fits is attested by both by the larger errors in the parameters and by
the larger values of $\chi^2$.  They obtained  profiles by
keeping  the position angle of the ellipses fixed, which introduces an error in the 
luminosity profile since   
isophotes are frequently  twisted.  Moreover, their 2D model considers  the apparent 
bulge and disk 
ellipticity to be fixed, while in the 1D model, both parameters were allow to vary.
We think that the result of these two analyses are not comparable. 

De Jong (1996) also compares the parameters of the bulge and disk fitted in the 1D and 2D
decompositions in his Figure 6. The only model used for all the 
galaxies in their 1D fit is that of
bulge+disk neglecting the obvious presence of bars in many of the galaxies of his sample;
 however,
the bars are taken into account in his 2D fit. Again, these two analysis cannot be compared.

\subsection{ The components }

      We assume that the luminosity distribution 
    of a galaxy is the sum of the distributions of its individual components. 
     We use the 
    following laws for fitting the different components found in the galaxy sample: 
    
\noindent {\it Bulges:} We approximate the bulge luminosity 
profile to  the $r^{1/n}$ law introduced by Sersic (1968):

\begin{equation}
I_{\rm bulge}(r)=I_{\rm e} 10^{-b_{n}((\frac{r}{r_{\rm e}})^{1/n}-1)},
\end{equation}
where $I_{\rm bulge}(r)$ is the surface brightness of the bulge in flux density at a distance  $r$ from the
 center,
or, in surface magnitude units:

\begin{equation}
\mu_{b}(r)=\mu_{\rm e}+c_{n}((\frac{r}{r_{\rm e}})^{1/n}-1),
\end{equation}
with $c_{n}=2.5b_{n}$, and where $b_{n}$ can be chosen in such a way that the
 scale-radius, $r_{\rm e}$, is the radius encircling half of the total luminosity, 
 $L_{\rm T}$, and $I_{\rm e}$ is the surface brightness at this radius.   

\noindent{\it Disks:}  The disk luminosity profiles were fitted by an exponential law, where the
surface brightness for the disk in flux density, $I_{\rm disk}(r)$, is given by:

\begin{equation}
I_{\rm disk}(r)=I_{od}e^{-r/h}.
\end{equation}
Or, in surface magnitude units, by,

\begin{equation}
\mu_{d}(r)=\mu_{o}+1.086 \frac{r}{h},
\end{equation}
where $I_{od}$ is the central surface brightness of the disk and $h$ is the scale length of 
the disk. 

\noindent{\it Bars:}  We find two types of bars.

\begin{enumerate}
\item {\em Elliptical bars}. These bars have elliptical isophotes and we have used the two-dimensional 
function given by Freeman (1966) 
for fitting the luminosity distribution, $I_{\rm bar}(x,y)$, of this type of
 bars:

\begin{equation}
I_{\rm bar}(x,y)=I_{o,{\rm bar}}\sqrt{1-(\frac{x}{a_{\rm bar}})^{2}-(\frac{y}{b}_{\rm bar})^{2}}
\end{equation}

where the free parameters are $I_{o,{\rm bar}}$, the bar central surface  
brightness, and $a_{\rm bar}$ and $b_{\rm bar}$,  which are the scale lengths for the 
semi-major and semi-minor 
axes, respectively, of the bar. This equation is valid within the bar region (i.e. $x\leq a_{\rm bar}$, $y\leq b_{\rm bar}$). In equation 5, the origin is the bar
center, the $x$-axis is in the direction of the bar major axis and
the $y$-axis is perpendicular to this.

\item {\em Flat bars.}  
 For fitting this kind of 
profile we use the expression given in Prieto {\em et al.} (1997): 

\begin{equation}
I_{\rm bar}(r)=\frac{I_{o,{\rm bar}}}{1+e^{\frac{r-\alpha}{\beta}}}, 
\end{equation}
where $\alpha$ and $\beta$ are constants which smooth the end of the bar.

\end{enumerate}

\noindent {\it Lenses:}  A lens is characterized by having  a smooth 
luminosity gradient
 with a very
 sharp cut-off.  Duval \& Athanassoula (1983), studying the galaxy NGC~5383, found 
  a lens in its luminosity profile and fitted it  to the expression:

\begin {equation}
I_{\rm lens}(r)=I_{ol}(1-(r/r_{ol})^{2}), r\leq r_{ol}
\end{equation}
where $I_{ol}$ is the surface brightness in flux density at the lens center , and $r_{ol}$ is
 the typical size of the 
lens.   We
 used this expression for fitting the lenses found in this sample.

\noindent {\it Rings:} Rings and pseudo-rings are present in many spiral galaxies.  
  The transverse luminosity profiles of these structures are well fitted with  
 Gaussian functions:

\begin{equation}
I_{\rm ring}(r)=I_{or}e^{-\frac{1}{2}(\frac{r-r_{or}}{\sigma})^{2}},
\end{equation}
where $I_{or}$ is the surface brightness in flux density at $r_{or}$, which is the radius of the ring,
 and $\sigma$ is the width   of the ring.

The total surface brightness model is the sum of all the above functions corresponding to the
components that appear in the galaxy:

\begin{equation}
I_{\rm total}(r)=I_{\rm bulge}(r)+I_{\rm disk}(r)+I_{\rm bar}(r)+I_{\rm lens}(r)+I_{\rm ring}+ ...
\end{equation}

\subsection{The fitting technique}

The most important  feature of the method is the  detailed study  of the  multicolor
photometric information for each of the galaxies, before 
 performing the
decomposition, to determine  
   the different structural components which
form the galaxies and  make an
 estimate of their scale lengths.  After the decomposition we   
 consider whether the  
 parameters obtained are physically meaningful. 
 
For an accurate decomposition, the intrinsic peculiarities of each galaxy 
do not permit an automatic method.
The non-linear $\chi ^2$ minimization routine
 method to fit the model profiles to the data points is quicker than  interactive
 methods, and perhaps  more suitable when treating a more extensive sample
 of galaxies,  but when we  analyze a small number of galaxies in
 detail, the automatic methods are not useful. For example, some points in the profiles of galaxies with a blue ring must be avoided 
 and the model must pass below them. An automatic method cannot do this.
  Moreover the $\chi ^2$  method is not useful for analyzing galaxies with
 complex structures. The  parameters obtained with this method are often not
physically meaningful.

  The process begins by examining the color and color-index images,
 which give us a qualitative 
  idea of what  the
 components are.  
  Next,
  we examine the color-index profiles, where we can estimate the scale length  of any additional components. Then, the confirmation of the assumed components is obtained in the ellipticity and position-angle profiles, where the geometry of the
 different structures  projected onto the sky plane are well reflected. 
 At this point we decide the set of structural components that make up the galaxy. 
 Then   we decompose the azimuthally averaged or individual (for galaxies with strong bars)
  luminosity profiles 
 of the galaxy to obtain the parameters of their components.

 We use  an interactive profile-fitting routine. 
 The routine begins by fitting the parameters for the disk and bulge over different
  ranges of the
 profile, defined by the linear trends of the surface brightness 
 against  $r$ and  $r^{1/n}$, respectively, with the
 least-squares method in an iterative process. Beginning with the  estimated initial values,
  we fit the other components (bars or lenses),  by varying their
  parameters until a good fit is obtained. These were then subtracted 
from the original
  profile and  the bulge and disk fitted again.
 This process is repeated until all parameters  convergence. We define  convergence to have occurred  when the difference between the 
 structural parameters of the bulge and
 disk for two consecutive steps is smaller than the fit errors of those parameters.

 For the galaxies, { NGC~1300}, { NGC~7479}, and { NGC~7723}, which have prominent bars, 
  we 
  used profiles along the major and minor
 axes of the bar instead of the azimuthal profile. On the minor-axis profile,
  (which is less affected
 by the bar component), corrected for
 inclination and position angle, we fit the bulge, disk, and  other components 
 using the method described above. 
 The fitted bulge and disk functions are subtracted from the major-axis profile to characterize the bar 
 in its long dimension. This process is repeated until all parameters convergence as described above.  
 Some bars, such as 
  {NGC~7723}, and {NGC~1300}, show  very strong star formation regions at their ends,  which were fitted  with  Gaussians. These
  regions are associated with the beginning of spiral arms.
The features created in a profile due to  spiral arm  crossing it  are also 
well fitted   
by Gaussian functions.

\section{Structural decomposition of the galaxies}

All the photometric information needed to analyze the galaxies is given in Paper I. The figures in that paper should be viewed in conjunction with this section.

The calibration constants in Paper I  include the correction due to  galactic
inclination and  absorption; consequently  all  parameters are
corrected for these effects. We have not corrected for internal extinction. According to
Xilouris {\em et al.} (1999), a typical face-on spiral galaxy is  transparent and is optically thick, at least in the central regions, down to inclination angles
of almost $60^\circ$. The galaxies in this sample are near to or below these limits. 
  
In  Figs. 1--11 we present the decomposition of the luminosity
 profiles in each filter and in  Tables I-XI  we show 
 the parameters of the models as defined in Sec. 2.1: the fractional luminosity of
 each component,  and the ellipticity and 
 position angle of the disk and bar defined as the average value of the last points
  for the disk and the value at the typical length for the bar.
  
  The uncertainty given for the bulge and disk parameters  comes from the
  ``standard errors'' of the  coefficients of the lines of regression of the
  fits. 
  These "standard errors" are a measure of the residuals between the observations and
  the fitted regression line. Clearly, these residuals depend  strongly  on the deformation
  of the profile due to  structures which were not fitted, such as spiral arms, star formations regions, etc... 
   
   The fit for bar and lens parameters was achieved by varying by a fixed amount the parameters of their
   luminosity laws around the estimated values. This number is the
   uncertainty in the parameters of the bars and lenses given in Tables 1--11.

     Next, we describe  the decomposition procedure for each galaxy. The morphological
  classification given is that of   de Vaucouleurs {\em et al.} (1991).

\noindent {\em NGC~1300.}  This is an SB(rs)bc galaxy.  In the color and color-index images and
 profiles (Fig.1a, and 2a, Paper I), we can distinguish a prominent bar  of about 70$''$
  (on the major axis),  a
very blue region at the center, 
  two dust lanes along the
bar, curved around the center, which could be related to the presence of an ILR (Athanossoula
1992a,b). Two prominent blue
spiral arms, and a uniform color region inside 50$''$ which suggests the presence 
of a lens. 
 The bar region has the same color as the disk, suggesting that the stars of the bar 
 are of the same type as those in the surroundings.  
 The galaxy seems to have  the following components:
 a    bulge, a disk, a bar,  and a lens. The ellipticity profile (Fig. 2a, Paper I)
  confirms the presence of these
      components. In Fig. 1  we present the profiles 
perpendicular and parallel to the bar with the various components fitted.
 The bulge is fitted with an index $n$=4 in all
filters, but not inside 10$''$ in the perpendicular bar profile, probably due to the 
strong extinction in this region.  
     The bar is very well fitted by an elliptical function.
    The model for the  parallel profiles inside 30$''$  does not 
    fit   the observations well, due to the two dust-lanes parallel to the bar.
     At the end of the bar there are star formation regions, which are well fitted by a Gaussian function. However, in the parallel profile at 150$''$ there is a hump in the luminosity due to a spiral arm
  which is not well fitted by a Gaussian, 
  due the presence in this direction of a very narrow peak, 
  probably caused by a  giant star formation region. 
  The hump  in the perpendicular profiles around 120$''$ is a
 spiral arm which is well fitted  by a Gaussian profile.  
  In this galaxy, the errors of the scale lengths of the disk are so large  that
  no conclusions can be  drawn concerning the relative values of the parameters of the different
  components or the bulge-to-disk (B/D) ratio.

\noindent {\em NGC~5992.} The morphological type of this galaxy is not clear; it is classified as
simply  S-type by de Vaucouleurs {\em et al.} (1991) and as Markarian 489 by Mazzarella \& Balzano (1986). It presents an active starburst 
nucleus (Balzano 1983; Bicay {\em et al.} 1995).
In the $B$$-$$I$ color-index image (Fig. 1b, Paper I)  some structure appears
  inside 10$''$, a red region to the NE 
from the center, and a blue one 
  to the E.
  In the ellipticity profile (Fig. 2b, Paper I) we observe a feature indicating a bar with a length of
  about $20\arcsec$. 
    We fit a luminosity 
  profile for the
bulge with an index  of 1.5 in all filters, 
an exponential disk, and a flat bar.   
The observational feature appearing above the model at about 6.5$''$  
 is probably due to the structure within $10\arcsec$.
   The B/D 
 ratio is high, 
which suggests that it is an early-type spiral  with structural deformations  
probably due to an interaction with NGC~5993. In Fig. 2 we present the fit of the all these parameters to
the surface-brightness profiles of the galaxies.

\noindent {\em NGC~6056:} This is a lenticular barred galaxy, classified  
as SB(s)0. 
  In the color-index map (Fig. 1c, Paper I) we  observe a structure of constant color,
  which probably corresponds to a
  lens ending at about $20\arcsec$.  The  ellipticity and position-angle  
 profiles reveal the presence of  a bar  about 8$''$ in length. 
 The lens is very faint in the optical profiles, but is prominent in the infrared filters, as we
 will show in a future paper.
The bar is flat and has a misalignment of about $10^\circ$ with respect to the line of nodes of the disk 
(see Fig. 2c, Paper I). 
   We obtained a very good fit  with a
    flat bar feature.  
    The $r_{\rm e}$ of the bulge increases, and the  scale length of the disk decreases, 
    with 
    redder filters. The scale length   of the bar in $I$ is smaller than in the other filters.

\noindent {\em NGC~6661.} This is  a lenticular galaxy, classified as SA(s)0/a.  
The color-index images and 
 profiles (Figs. 1d, and 2d, Paper I) present a constant-color 
region from about 10$''$ to 30$''$, which suggests the presence of a lens. A red ring  appears
 at about 30$''$ in the color-index profiles and $B$$-$$I$ image.
  We need to add an elliptical lens to the bulge and disk functions for a good fit to the 
  luminosity profiles in all filters (Fig. 4).
 The bulge is fitted with an index of
 about 2  in all filters, the disk is smaller 
 in redder filters, and the  scale length of the lens is smaller in   $I$ than 
 in the other filters.

\noindent {\em NGC~6946.}  This is a late-type spiral classified as SAB(rs)cd, with an active nucleus,
  type HII (Keel 1984; Engelbracht {\em et al.} 1996). In the $B$ and $U$ filters it was not possible to fit ellipses to the
  isophotes due to the
 presence of strong spiral arms and many high luminosity regions, probably
  star
 formation regions. The color and color-index 
  images (Fig. 1e, and 2e, Paper I) show an extensive central region of constant  color, suggesting the presence of a lens. In the color index  profile, we can estimate the length
  scale of this component, about $70''$, and this can be confirmed in the ellipticity 
  profiles where there is a change in the ellipticity at just about
 this distance.
 The misalignment between the inner and the outer isophotes seen in the 
position-angle
  profiles
 (Fig. 2e, Paper I)
   may  indicate the presence of a triaxial bulge.
 
 The fluctuations of the observations about the disk model (Fig. 6) are due to the presence of spiral arms; there is a very blue arm at about 20$''$, which  causes the model to fall 
  below the observations in the bluer filters. The 
 bulge  fits   an $r^{1/4}$ law, including in the central part, and its
 relative luminosity increases toward the redder filters.
 The scale length  and relative luminosity of the disk decreases greatly with the redder filters. 
 This galaxy presents the smallest B/D ratio of the sample.

\noindent {\em NGC~7013.} This is another lenticular galaxy, classified as SA(r)0/a.  
 Optical and H$_\alpha$ images (Lynds 1974) show a small bulge, and an inner ring. The H~{\sc i}
distribution is in two rings (Eskridge \& Pogge 1991; Knapp {\em et al.} 1984). The larger ring is
situated just inside the edge of the optical disk, and the smaller one is associated with the inner stellar
 ring. In the  color-index images and profiles (Figs. 1g, and 2g, Paper I) we  observe a blue
 ring feature 
centered at about 25$''$, and  a constant-color region from 10$''$ to 
60$''$, suggesting 
 the presence of a lens. In the ellipticity
    profile we can confirm these components. At about
    25$''$ the isophotes have an ellipticity close to 0.65 corresponding to the position of the 
     ring; this is   similar to that of the outer disk. After
    this, there are a plateau until about 60$''$, which corresponds to the lens.
     This lens probably has a non-zero intrinsic ellipticity because  it is considerably smaller than that
    of the outer disk, and its position angles are quite different.   
    Beyond this, the ellipticity and position angle of
    the ellipses are due to the inclination of the galactic disk.
    We have modeled
    the observed luminosity profiles of this galaxy with four components (Fig. 6): a bulge, a disk, a
     lens, and a
    ring. The value of the index $n$ for the bulge depends on the filter, and is about 1.5. 
    The  scale length for the disk and lens is smaller  in $I$ than in  the other filters.

\noindent {\em NGC~7217.} This galaxy is classified as (R)SA(r)ab.  It is a LINER (Ho, Filippenko, \& Sargent 1993;
  Hummel {\em et al.} 1987). In  the $B$$-$$I$  and $B$$-$$V$ color map and profiles (Figs. 1g, and 2g, Paper I)
   we can see  
   the ring structures, with three nuclear rings  (red,  blue and 
 another red) at about  8$''$,
   10$''$ and 15$''$,   a 
   blue inner ring,
and a blue outer ring at about 30$''$, and 75$''$, 
respectively, measured in the $B$$-$$I$ image. The
ellipticity and PA profiles indicate
 that the rings are quite circular, and
that their isophotes 
have similar position angles to those of the disk. Buta {\em et al.} (1995) and  Verdes-Montenegro, Bosma,
 \& Athanassoula (1995) studied the ring
structures of this galaxy and their
 locations are in agreement with our values.   
The bulge is fitted with an  $r^{1/2}$ law in all filters. In addition to the bulge and disk
 in the $B$ and $V$ filters, we have fitted the blue nuclear
ring, and the blue inner and outer rings; in the $R$ and $I$ filters we 
have also fitted the red  innermost
 ring and the
blue outer ring.  All rings are fitted
 with Gaussian functions.
   In Table 7, we give the  positions of the rings in the different filters from the
 model fit, as shown in Figure 7.

\noindent {\em NGC~7479.} This is a barred galaxy, cataloged as SB(s)bc;  
it is classified as a LINER type by Keel (1983a,b) and Devereux (1989), and as HII (Hummel {\em et
al.} 1987). This galaxy has been studied by many authors. Dynamic studies  by
Laine (1996) reveal the presence of an interaction with another galaxy, which could explain the
asymmetry of the arms. It presents large, bright H~{\sc ii} 
regions along the bar (Hua, Donas, \& Duun 1980). The  central blue
peak that appears in the $B$$-$$I$ color-index profile is
probably  due to  nuclear activity. 
This galaxy shows a
prominent asymmetric bar,   and strong spiral arms in the color and color-index
 images (Figs. 1h, and 2h, Paper I). In the bar region there
are strong dust lanes, which present an asymmetric distribution. The color of the bar and
 bulge is redder than that of the disk. The
ellipticity and position angle profiles (Fig. 2h, Paper I) have the features of a strong bar  
 about 100$''$ in semi-length, its position angle being  about $100^\circ$,  different from that of
the disk. The presence of a triaxial bulge is suggested by the misalignment 
between the inner
and outer isophotes.  
  The bar is  fitted with a flat
  function along the both axes  (Fig. 8).
    In these profiles, it is possible to see two 
  spiral-arm features,  at around $40\arcsec$ and $100\arcsec$. 
   This inner spiral arm is very smooth in the $I$ band (it is almost 
  invisible) but  is clearly  present in the $V$ band.  Elmegreen \& 
  Elmegreen (1985), studying the bar--interbar intensities,  also found that this
   bar has a flat light profile.  The fit
    is better in the $I$ band than in $R$ and $ V$,   
     due the $I$ profile is less affected by the star formation regions present along the bar's
     major axis (Hummel {\em et al.} 1987).
      The bulge profile fits  an exponential function well. As in NGC~1300, the large
     error in the scale length  of the disk prevents us from reaching any conclusions concerning the trend of
     the parameters with the filters.

\noindent {\em NGC~7606.} This galaxy is cataloged as SA(s)b with a LINER-type nucleus (Keel 1983b). 
 The color and color-index 
  images (Figs. 1i, and 2i, Paper I) show a red nucleus, a constant-color region inside 20$''$,  and beyond this there 
  are very prominent 
spiral arms. 
 In the $B$$-$$I$ map there are  three red pseudo-rings, probably due to  the inter regions. 
 The ellipticity and position-angle profiles (Figs. 1i, and 2i, Paper I) confirm these structures; inside 
 20$''$ the bulge geometry dominates. From here out to about 60$''$ a
 different structure appears, due to the arms and rings,  
  and beyond 
 this distance the ellipticity and position angle are constant due the inclination of the disk. 
 The bulge+disk model provides a good fit to the
observations, including the central part of the bulge, with
an  index $n$ of around  2  for all filters.  Andredakis \& Sanders (1994) also  found an
exponential bulge (Fig. 9).
 The early
type of this galaxy, together with its high inclination, cause 
 the
strong spiral arms to look like rings; they are prominent in the isophotal
 luminosity profiles above the bulge-plus-disk model.

\noindent {\em NGC~7723.} This is an SB(r)b   galaxy  with HII-type nuclear activity (Keel 
1983b; Hummel {\em et al.} 1987; Giurcin {\em et al.} 1994). 
In the color and color-index maps and profiles (Figs. 1j, and 2j, Paper I) 
there is clear evidence for  a bar 
 of semi-length about
 30$''$  and  a pseudo-ring at the end of the bar. The bar is redder than the disk 
and  shows
two straight dust-lanes emanating from the nucleus. The structure of these dust-lanes, which are not curved inward toward the center, suggests that 
there is no
 ILR in this galaxy (Athanassoula 1992a,b). The central pixels of this galaxy are 
saturated
 and the results are not significant inside the central 5\arcsec.  The color index 
of the disk is quite constant with radius. In the ellipticity and position-angle 
profiles (Fig. 2j, Paper I)
 the bar and pseudo-ring are clearly present, with  different position angles from that of the disk. The bar is well fit  by an 
  elliptical bar function,  and the pseudo-ring   by a Gaussian function.  The hump that 
appears in the perpendicular profiles around 30$''$ corresponds to the 
pseudo-ring. 
  This galaxy has  smooth and very broad spiral arms, 
 which correspond to the region (from 50$''$ to 90$''$), where the model does not fit 
  the observed profiles. The bulge is fitted with an index $n$=1 ( exponential law).
   The $r_{\rm e}$ for the bulge and the $h$ for the disk are constant in all the  bands.

\noindent {\em NGC~7753.} It is in interaction with another galaxy (Salo \& Laurikainen 1993)  and  is classified as SAB(rs)bc. 
  In the color and 
  color-index maps (Figs. 1k, and 2k, Paper I) a small (about 10$''$ 
  long) bar is evident. The existence of  
such structure is also indicated by the ellipticity and position-angle profiles (Fig. 2k, Paper I). The 
luminosity profiles
were modeled with a bulge, a disk, and an elliptical bar (Fig. 11).
  At around 40$''$ there is an increase in luminosity above the model which is due to a
spiral arm.
 The bulge was fitted with an $r^{1/2}$ law.

 \section{Discussion}

 To  analyze the results of the structural decomposition of this sample of galaxies, we 
  divide these in three groups: those with a well defined disk, those with 
an ill-defined disk and those with strong bars
 and well developed arms.
 
 1.- Galaxies with a well defined disk. 
 
 These galaxies have an extensive part of their disk free from contamination by 
  other components, which implies that the fitted 
 bulge and disk parameters are reliable, and that we can make
 comparisons between
 them. Our criterion for selecting these galaxies is that they have at least $1/3$ of their
 luminosity  profile
 coming from the disk without overlapping  other components. Only four galaxies obey this condition:
 NGC~6056, NGC~6661, NGC~6946, and NGC~7606 (NGC~6661 fault this condition in the $I$ band.
 
 The following interpretations, although of an outstandingly coherent behavior, are made 
  with caution due the small number of galaxies and the reduced statistical significance.
 
 In Fig. 12 we present the shape index, $n$, versus the filter.
   There is a trend where $n$  increases with the redder filters. This can be
   interpreted as the older population being more extended than the younger,
which could indicate an 
   end of the star formation in
   the bulge from outside to inside.  Extinction 
    would have the reverse trend. The absolute variation of  $n$ with  filter is 
 $\pm0.3$ around the mean value. The $n$ index is bigger for the Sb galaxy than for the Scd one.
  We are not going to compare this index between spirals and lenticulars.

 \begin{figure}[!htb]
    \begin{flushleft}
\epsfig{figure=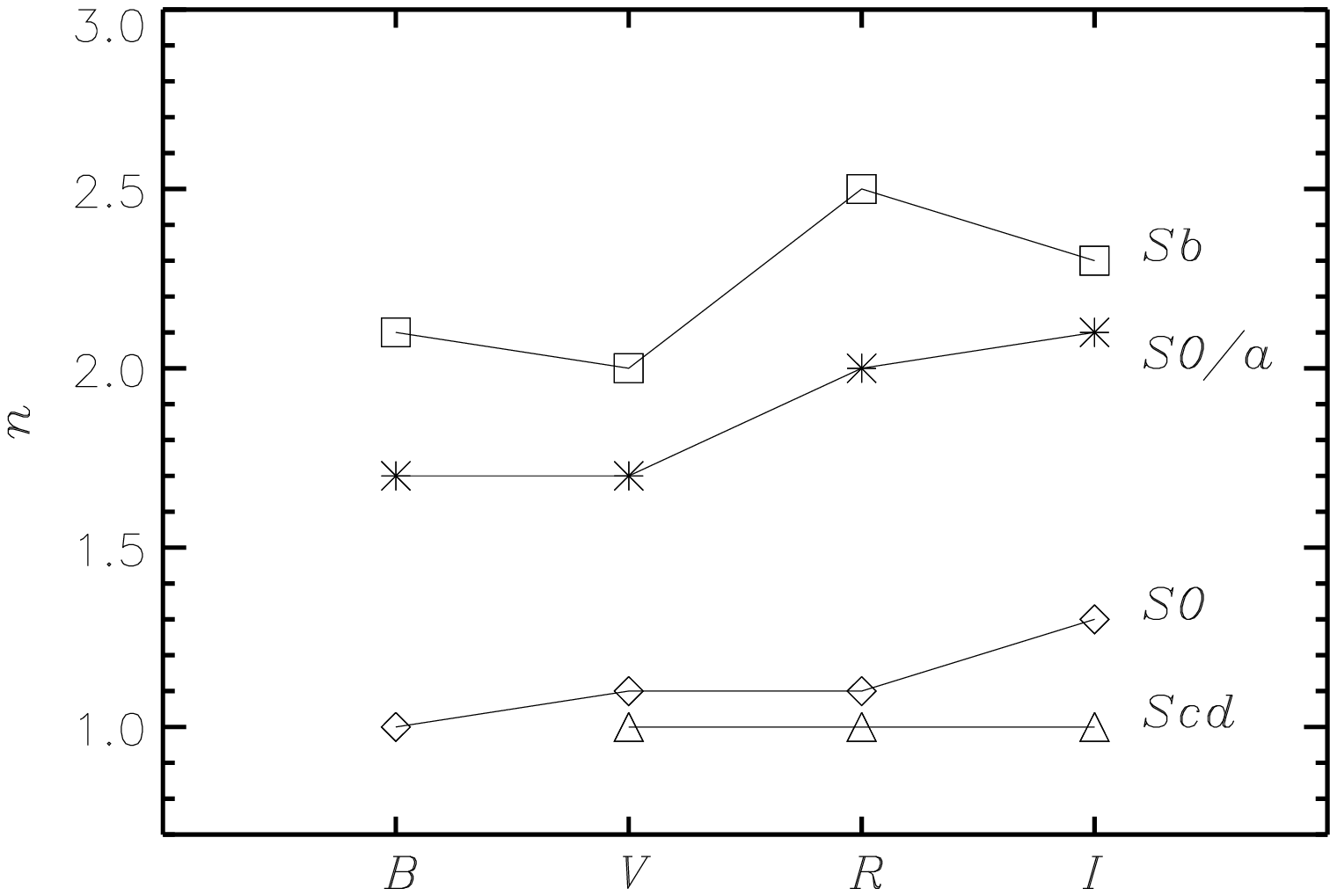,height=9.truecm,width=9.truecm}
            {\bf{Fig 12.}}  { The index $n$ of the bulge model vs  filter for those galaxies with
well defined disk and variable index.}
    \end{flushleft}
    \end{figure}

 These  galaxies are of different morphological types (SO, SO/a, Sb, and Sc). Their 
 inclinations are the largest of the sample, near the transparent limits  
 ($60^\circ$),
  (Xilouris {\em et al.} (1999)), so we expect to find some effect of the internal extinction
   on the scale length of the disk, $h$,  depending on the filter. In Fig. 13 we see  how $h$  varies with the filter for each galaxy. The scale length of the disk
    decreases in all
   galaxies when the filter is redder, as we would expect an optical
 thickness increasing
   towards the center of the galaxy. However, we can not exclude an additional effect due to stellar
   population.
   The effective radii of the bulges do not present any systematic trend with the
   filter.

 \begin{figure}[!htb]
    \begin{flushleft}
\epsfig{figure=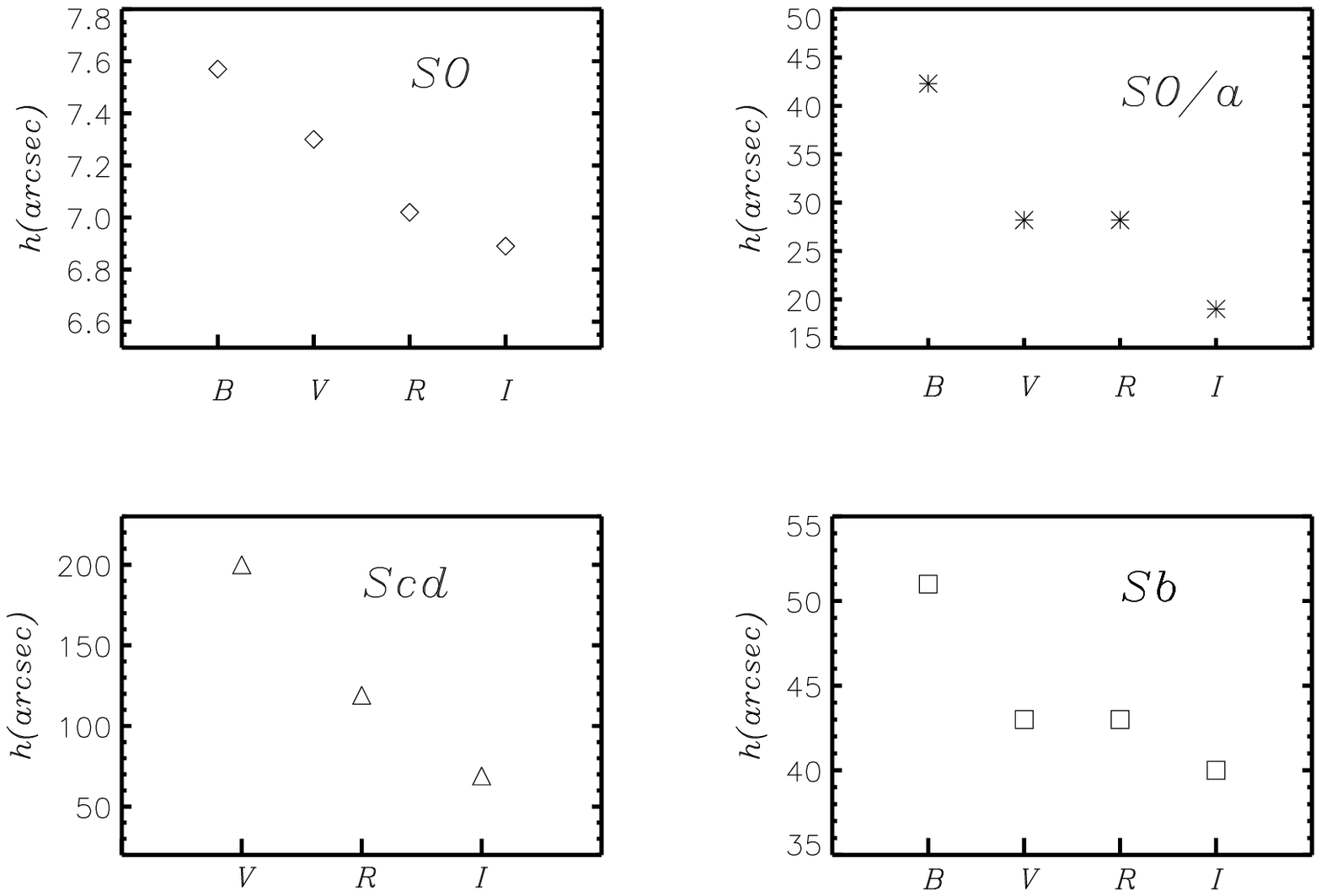,height=9.truecm,width=9.truecm}
             {\bf{Fig 13.}} { Scale length of the disk against  filter for those galaxies with well
defined disk.}
    \end{flushleft}
    \end{figure}

   The ratio of the scale length of the disk to the effective radius of the bulge,
   Fig. 14, presents a systematic decrease to the red. We also find a systematic increase of
    $r_{\rm e}/{\rm h}$ from 
   late to early type galaxies, contrary to that found by Courteau,
de Jong, \& Broeils (1996) and de Jong
   (1996).  This problem have been rigorously treated in Graham \& Prieto (1999).

   2.- Galaxies with an ill-defined disk.
   
   Although in these galaxies the disk is not well defined, we have carried out the
   structural decomposition in order to determine which  components are present and we have obtained an estimate of their scale lengths. 
   
   {\em NGC~7013}. This galaxy has a blue outer ring,  seen clearly in the $B$$-$$V$ and $B$$-$$I$ images
   and in the luminosity profiles. This ring hinders the clear detection of
 the disk and  this has
   been fitted to the external edges of the ring. The galaxy has
    an inclination close to the transparency limit and   there is a trend toward decreasing $h$ in redder filters. We can conclude that this galaxy
   has a ring at 20$''$, a lens of scale 
length is 50$''$, and a 
bulge shape index of 1.5.
   
   {\em NGC~7217}. This galaxy has multiple blue and red rings which hinder 
clear detection of
   the disk. We have fitted the disk to the zone we consider to be between
   rings, but these rings are very wide  and numerous  except for the $I$ filter where they
    are smoother. 
   
   {\em NGC~5992}. This galaxy is probably interacting with NGC~5993. Except in the $B$ filter, there is too
   little data for the outer disk to make a good estimate of the disk parameters. We have fitted the disk to the outermost points
   of the galaxy, outside the bar region. We can conclude that
    this galaxy has a flat bar with scale length is 19$''$.
   
   {\em NGC~7753}. This galaxy has very strong spiral arms. We have carry out the fit of the
   disk in the inter-arm regions. 

 \begin{figure}[!htb]
    \begin{flushleft}
\epsfig{figure=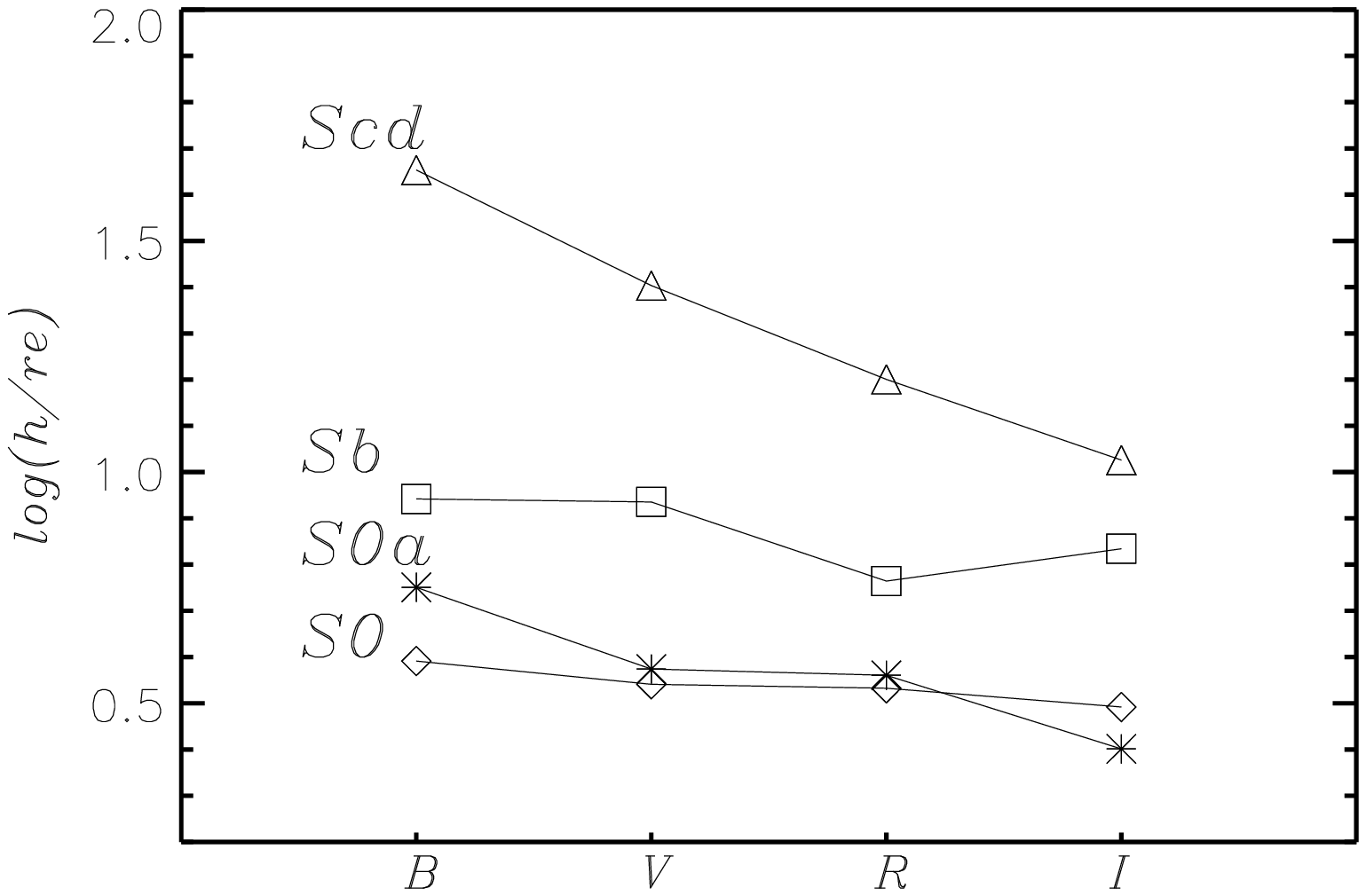,height=9.truecm,width=9.truecm}
              {\bf{Fig 14.}} { The ratio $h/r_{\rm e}$ vs the filter for those galaxies with well
defined disk.}
    \end{flushleft}
    \end{figure}

   3.- Galaxies with large bars and well defined spiral arms.
   
    These galaxies are, NGC~1300, NGC~7479, and NGC~7723. None of these galaxies have a  well observed
    region of the disk due to the presence of powerful spiral arms. We
    have fitted the disk in individual profiles  and
    on the existing points between the arms. 
There is no clear zone of disk in any of these galaxies. 
However, we have achieved the fittings to know the general features of
    the other components.  
    
    {\em NGC~1300}. This galaxy has  a bar, a lens, prominent spiral arms and star 
    formation
    regions.
    
    {\em NGC~7479 and NGC~7723}. These galaxies have very few points to fit the disk.
    They have an elliptical bar with scale length of 51$''$ and scale wide
    25$''$ in NGC~7479 and scale length of 22$''$ and scale wide of 12$''$ for NGC~7723.

  The bulges of the spiral galaxies in this sample, with the exception of NGC~1300, have an $n$ index
  which follows the general trend observed in spiral galaxies. The earliest, NGC~7217
  (Sab), has an index $n=2.8$, and the rest have values between 1 and 2.

We did not find Freeman type II profiles in this sample of 11 galaxies probably because we considered 
components other than  the bulge and disk. 

We found bars in  half  of the   galaxies analyzed.
Half of these bars have the same color as the underlying structure (NGC~1300, NGC~6056,
and NGC~5992), and the other half are redder than such structure (NGC~7479, NGC~7723, and NGC~7753). 
The cause of this segregation is a question still to be understood.  This could 
give us some indication of the state of the bar; the red bars are probably
the oldest, and the young bars retain
the color of the disk stars which give them their form.

\section{ Summary and Conclusions}

We have performed a structural decomposition for a sample of 11 disk galaxies of
different morphological types. 
 The bulges were fitted with an $r^{1/n}$ law, the disks with an
 exponential law, the bars with elliptical or flat functions, rings and spiral
 arms with Gaussian functions, and the lenses with a quadratic expression.
 Prior to the fit, we  used the $U$, $B$, $V$, $R$, and $I$ photometric data
 (color and color-index images and luminosity, 
ellipticity, and position-angle profiles) to decide
 the type and number of  different components which form the galaxies
and to estimate their scale lengths. We have written an interactive profile-fitting routine for the decomposition, which fits the parameters of the models in 
an iterative process. We find and model all components which form the galaxies: 
bulge, disk, bar, lens, ring, etc. Only for galaxies with well defined disks
do we give reliable
parameters for the bulge and disk.

For the galaxies with well defined disks we find that : 

 --  The scale length of the disk
    decreases  when the filter is redder, as  would be expected for an optical thickness 
    increasing towards the center of the galaxy. 

 -- There is a  increased trend in the index $n$ of the bulge
with the redder filters.
  This can be
   interpreted as the older population being more extended than the younger.

-- The ratio between the scale length of the disk and the effective radius of the bulge,
   shows a systematic decrease to the red.
    We  also find a systematic increase in this ratio from  late 
    to early types of galaxies.
   
   For the galaxies with ill-defined disk we have given the position and length-scales of 
   components other than the bulge and disk.

We do not find Freeman type II profiles in this sample of galaxies, probably because  we considered 
components other than  the bulge and disk. 

We found bars in  half  of the   galaxies analyzed, which are either
elliptical or flat.
Half of these bars have the same color as the underlying structure (NGC~1300, NGC~6056,
and NGC~5992), and the other half have redder colors (NGC~7479, NGC~7723, and NGC~7753).


 We are most sincerely grateful to Alister Graham  for useful conversations and  kindly reading 
 and improving
 the manuscript. We express 
our thanks to  Terry Mahoney for correcting the English of the
manuscript. The 2.5-m INT is operated on the island of La Palma by the 
Royal Greenwich Observatory at the Spanish {\em Observatorio del Roque 
de Los Muchachos} of the {\em Instituto de Astrof\'\i sica de Canarias}.
 Support for this work comes from project PB97-1107 and PB97-0219 of the Spanish DGES.
  The observations    received financial support from the European
   Commission 
through the Access to Large-Scale Facilities Activity of the Human 
Capital and Mobility Programme.

\clearpage

 \begin{table*}
	\begin{center}
	{\bf{Table 1.}} Structural parameters of NGC 1300.
	\end{center}
      \[
	 \begin {tabular}{ccccc}
            \hline
	Structure & Parameters & $B$ & $R$ & $I$  \\ \hline 
	Bulge &   $r_{e}$(") & $2.86 \pm 0.17$ & $5.58 \pm 0.15$ & $3.9 \pm 0.3$ \\  
        & $\mu_{e}$ (mag/$arcsec^{2}$) & $18.45 \pm 0.16$ & $18.5 \pm 0.1$ & $17.2 \pm 0.2$\\ 
	& n & 4.0 & 4.0  & 4.0 \\ 
	& \% $L_{Bulge}/L_{Total}$ & $30\pm 1$ & $29\pm 1$ & $19\pm 1$ \\ \hline
	{ Disk} & $h$(") & $58 \pm 21$ & $70 \pm 10$ & $101 \pm 25$ \\ 
	& $\mu_{0}$(mag/$arcsec^{2}$) & $23 \pm 0.8$ & $21.9 \pm 0.5$ & $21.5 \pm 0.8$\\

	& Ellipticity  & $0.39 \pm 0.11$ & $0.45 \pm 0.07$ & $0.42 \pm 0.08$ \\  
	& Position Angle  & $96 \pm 10$& $97 \pm 7$ & $94 \pm 8$ \\
	& \% $L_{Disk}/L_{Total}$ & $52\pm 30$ & $55\pm20$ & $68\pm29$ \\
	& B/D & $0.5\pm0.1$ & $0.5\pm0.1$  & $0.3\pm0.1$  \\ \hline
	{Bar} & Major axis (Type) & Elliptical & Elliptical & Elliptical\\ 
	& $\mu_{0}$(mag/$arcsec^{2}$) & $23.61\pm0.07$ & $22.61\pm0.03$ & $21.52\pm0.02$\\ 
	& $a (")$ & $95\pm1$ & $87\pm1$ & $85\pm1$ \\ 
 	& Minor axis (Type) & Elliptical & Elliptical & Elliptical \\ 
	& $b (")$ & $32.4\pm0.5$ & $43.2\pm0.5$ & $416\pm0.5$ \\ 
	& Position angle & $ 100 \pm 4 $& $ 101\pm2 $& $101 \pm 2 $ \\ 
	& Ellipticity (b/a) & $0.34\pm0.01$&$0.49\pm0.01$ & $0.43\pm0.01$\\ 
	& \% $L_{Bar}/L_{Total}$ & $9\pm3$ & $7.4\pm0.2$ & $7.67\pm0.05$ \\ \hline
	Lens & $\mu_{l0}$(mag/$arcsec^{2}$) & $23.75\pm0.07$ & $22.75\pm0.03$ & $21.83\pm0.03$ \\ 
	& $r_{ol}$ & $64.8\pm0.5$ & $78.3\pm0.5$ & $67.5\pm0.5$ \\ 
	& Position angle & $100 \pm 4$& $100 \pm 4$ & $100 \pm 3$\\ 
	& \% $L_{Lens}/L_{Total}$ & $8.15\pm0.03$ & $7.97\pm0.14$ & $5.58\pm0.05$ \\ \hline
	\end{tabular}
      \]
   \end{table*}

\begin{table*}
	\begin{center}
      {\bf{Table 2.}} Structural parameters of NGC 5992.
	\end{center}
      \[
	 \begin {tabular}{cccccc}
            \hline
	{Structure} & {Parameters} & { B} & { V}& { R} & { I}  \\ \hline 
	{Bulge} &   $r_{e}$(") & $4.24\pm0.06$ & $3.65\pm0.08$ & $3.85\pm0.05$  & $3.91\pm0.09$ \\   
        & $\mu_{e}$ (mag/$arsec^{2}$) & $20.27\pm0.02$ & $19.78\pm0.03$ & $19.28\pm0.02$ & $19.14\pm0.02$ \\ 
	& n & 1.5 & 1.5  & 1.5 & 1.5  \\ 
	& $\% L_{Bulge}/L_{Total}$ &$41.2\pm0.6$ & $36\pm5$&$45.6\pm0.6$ &$47\pm2$\\  \hline
	{Disk} & $h$(") & $8.63\pm0.27$ & $8.18\pm0.14$ & $8.85\pm0.25$ & $11.15\pm0.04$ \\ 
	& $\mu_{0}$(mag/$arsec^{2}$) & $20.75\pm0.27$ & $20.3\pm0.1$ & $20.26\pm0.07$ & $20.76\pm0.04$ \\ 

	& Ellipticity  & $0.22\pm0.12$ & $0.23\pm0.15$ & $0.21\pm0.12$ &  $0.21\pm0.15$ \\ 
	& Position Angle  & $26\pm18$& $26\pm23$ & $39\pm18$ & $36\pm35$  \\  
	& $\% L_{Disk}/L_{Total}$ & $47.98\pm0.34$& $51\pm4$& $42.8\pm0.6$ & $38\pm2$ \\ 

	& B/D & $0.85\pm0.05$ & $0.75\pm0.06$  & $1.06\pm0.01$ & $1.25\pm0.07$   \\ \hline
	{Bar} & Type & Flat & Flat & Flat &Flat  \\ 
	& $\mu_{0}$(mag/$arsec^{2}$) & $23.3\pm0.1$ & $22.8\pm0.1$ & $22.5\pm0.1$ & $22.1\pm0.1$ \\ 
	& $\alpha (")$ & $19.44\pm0.5$ & $18.9\pm0.5$ & $18.9\pm0.5$ & $18.9\pm0.5$  \\ 
	& $\beta (")$ & $1.1\pm0.5$ & $1.1\pm0.5$ & $1.1\pm0.5$ & $1.1\pm0.5$ \\ 
	& Position Angle & $89\pm3$&$93\pm3$ &$93\pm3$ &$93\pm3$    \\ 
	& Ellipticity  &$0.49\pm0.04$ &$0.51\pm0.04$ & $0.50\pm0.03$ &  $0.51\pm0.04$\\  
	& $\% L_{Bar}/L_{Total}$ & $11\pm5$& $13\pm6$&$12\pm6$ & $15\pm6$  \\ \hline  
	\end{tabular}
      \]
   \end{table*}

\begin{table*}
      \begin{center}
	{\bf{Table 3.} Structural parameters of NGC 6056.}
	\end{center}
      \[
	 \begin {tabular}{cccccc}
            \hline
	{Structure} & {Parameters} & {B} & { V}& { R} & { I}  \\ \hline 
	{Bulge} &   $r_{e}$(") & $1.94\pm0.02$ & $2.1\pm0.2$ & $2.06\pm0.07$  & $2.22\pm0.13$ \\
        & $\mu_{e}$ (mag/$arcsec^{2}$) & $21.65\pm0.02$ & $20.74\pm0.02$ & $20.15\pm0.02$ & $19.70\pm0.06$ \\
	& n & 1.0 & 1.1  & 1.1 & 1.3 \\
	& \% $L_{Bulge}/L_{Total}$ & $7.54\pm0.02$ & $9\pm2$ & $9.7\pm0.5$ & $9.10\pm0.06$ \\ \hline
	{Disk} & $h$(") & $7.57\pm0.07$ & $7.30\pm0.12$ & $7.02\pm0.05$ & $6.89\pm0.15$ \\
	& $\mu_{0}$(mag/$arcsec^{2}$) & $21.29\pm0.02$ & $20.34\pm0.06$ & $19.67\pm0.07$ & $18.93\pm0.08$ \\

	& Ellipticity  & $0.45\pm0.06$ & $0.47\pm0.03$ & $0.44\pm0.06$ & $0.42\pm0.16$ \\
	& Position Angle  & $53\pm5$& $53\pm6$ & $53\pm6$ & $54\pm15$ \\
	& \% $L_{Disk}/L_{Total}$ & $83.8\pm0.4$ & $83\pm3$ & $88\pm8$ & $84\pm5$ \\

	& B/D & $0.10\pm0.01$ & $0.11\pm0.02$  & $0.11\pm0.01$ & $0.11\pm0.01$  \\ \hline
	{Bar} & Type & Flat & Flat & Flat &Flat  \\ 
	& $\mu_{0}$(mag/$arcsec^{2}$) & $23.17\pm0.03$ & $22.53\pm0.02$ & $21.92\pm0.02$ & $21.09\pm0.02$ \\ 
	& $\alpha (")$ & $8.6\pm0.5$ & $9.2\pm0.5$ & $8.6\pm0.5$ & $8.1\pm0.5$  \\ 
	& $\beta (")$ & $1.1\pm0.5$ & $1.1\pm0.5$ & $1.1\pm0.5$ & $1.1\pm0.5$ \\ 
	& Position Angle & $ 65\pm1 $& $ 67.7\pm0.5 $& $67.5\pm0.5 $ &  $ 65.4\pm0.7$   \\ 
	& Ellipticity  & $0.48\pm0.01$&$0.52\pm0.01$ & $0.51\pm0.01$ & $0.51\pm0.01$ \\ 
	& \% $L_{Bar}/L_{Total}$ & $8.62\pm0.06$ & $8.12\pm0.08$ & $7.45\pm0.03$ & $7.40\pm0.03$ \\ \hline

	\end{tabular}
      \]
   \end{table*}

\begin{table*}
      \begin{center}
{\bf{Table 4.} Structural parameters of NGC 6661.}
	\end{center}
      \[
	 \begin {tabular}{cccccc}
            \hline
	{Structure} & {Parameters} & { B} & { V}& { R} & { I}  \\ \hline 
	{Bulge} &   $r_{e}$(") & $7.51\pm0.07$ & $7.52\pm0.09$ & $7.56\pm0.07$  & $7.54\pm0.03$ \\   
        & $\mu_{e}$ (mag/$arsec^{2}$) & $21.13\pm0.21$ & $20.15\pm0.01$ & $19.67\pm0.01$ & $19.40\pm0.01$ \\ 
	& n & 1.7 & 1.7  & 2.0 & 2.1  \\ 
	& \% $L_{Bulge}/L_{Total}$ & $17\pm3$ & $27.8\pm0.5$ & $35.8\pm0.4$ & $32.8\pm0.2$ \\ \hline

	{Disk} & $h$(") & $42\pm2$ & $28\pm1$ & $28\pm1$ & $19\pm1$ \\ 
	& $\mu_{0}$(mag/$arsec^{2}$) & $21.81\pm0.06$ & $21.0\pm0.1$ & $21.0\pm0.1$ & $19.62\pm0.14$ \\ 

	& Ellipticity  & $0.46\pm0.12$ & $0.34\pm0.14$ & $0.30\pm0.14$ & $0.3\pm0.1$  \\ 
	& Position Angle  & $60\pm10$& $56\pm14$ & $53\pm16$ & $51\pm12$  \\ 
	& \% $L_{Disk}/L_{Total}$ & $76\pm6$ & $63\pm1$ & $56\pm2$ & $64\pm6$ \\

	& B/D & $0.22\pm0.05$ & $0.44\pm0.01$  & $0.64\pm0.01$ & $0.51\pm0.01$   \\  \hline
	{Lens} & $\mu_{l0}$(mag/$arsec^{2}$) &$21.62\pm0.04$  & $21.33\pm0.02$ &$21.21\pm0.04$ &$21.14\pm0.05$  \\ 	
& $r_{ol}$ & $23.7\pm0.5$ & $24.3\pm0.5$ & $23.2\pm0.5$& $17.8\pm0.5$\\ 
	& Position Angle & $53\pm6$&$46\pm9$  & $49\pm3$ & $47\pm2$ \\
	& \% $L_{Lens}/L_{Total}$ & $7.1\pm0.1$ & $8.9\pm0.1$ & $7.91\pm0.03$ & $3.58\pm0.01$ \\
 \hline
	\end{tabular}
      \]
   \end{table*}

\begin{table*}
      \begin{center}
	{\bf{Table 5.} Structural parameters of NGC 6946.}
	\end{center}
      \[
	 \begin {tabular}{ccccc}
            \hline
	{Structure} & {Parameters}  & { V}& { R} & { I}  \\ \hline 
	{Bulge} &   $r_{e}$(") & $3.73\pm0.01$ & $3.62\pm0.02$ & $3.5\pm0.1$ \\    
        & $\mu_{e}$ (mag/$arcsec^{2}$) & $17.81\pm0.07$ & $17.32\pm0.04$ & $16.78\pm0.05$ \\ 
	& n & 1.0 & 1.0  & 1.0   \\
	& \% $L_{Bulge}/L_{Total}$ & $0.61\pm0.03$ & $0.71\pm0.12$ & $3\pm2$ \\ \hline

	{Disk} & $h$(") & $174\pm21$ & $114\pm7$ & $68\pm2$   \\ 
	& $\mu_{0}$(mag/$arcsec^{2}$) & $19.94\pm0.06$ & $19.29\pm0.05$ & $18.73\pm0.03$  \\ 

	& Ellipticity  & $0.67\pm0.11$ & $0.63\pm0.06$ & $0.58\pm0.11$   \\ 
	& Position angle  & $101\pm7$& $119\pm5$ & $131\pm8$   \\
	& \% $L_{Disk}/L_{Total}$ & $98\pm1$ & $98\pm1$ & $95\pm2$ \\ 

	& B/D & $0.006\pm0.002$ & $0.011\pm0.001$  & $0.030\pm0.001$    \\ \hline 
	{Lens} & $\mu_{l0}$(mag/$arcsec^{2}$) & $19.8\pm0.1$ & $19.6\pm0.1$ & $19.3\pm0.1$ \\  
	& $r_{ol}$ & $27.5\pm0.5$ & $27.5\pm0.5$ & $27.5\pm0.5$ \\ 
	& Position angle & $110 \pm 4$& $109 \pm 4$ & $109 \pm 3$\\ 
	& \% $L_{Lens}/L_{Total}$ & $0.67\pm0.04$ & $0.64\pm0.03$ & $2.2\pm0.4$ \\ \hline
	\end{tabular}
      \]
   \end{table*}

\begin{table*}
      \begin{center}
{\bf{Table 6.} Structural parameters of NGC 7013.}
	\end{center}
      \[
	 \begin {tabular}{cccccc}
            \hline
	{Structure} & {Parameters} & { B} & { V}& { R} & { I}  \\ \hline 
	{Bulge} &   $r_{e}$(") & $7.37\pm0.15$ & $7.56\pm0.13$ & $7.74\pm0.1$ &$6.41\pm0.10$  \\   
        & $\mu_{e}$ (mag/$arcsec^{2}$) & $20.14\pm0.02$ & $19.10\pm0.03$ & $19.10\pm0.01$ &$18.40\pm0.04$ \\ 
	& n & 1.3 & 1.5  & 1.6&1.5  \\ 
	& \% $L_{Bulge}/L_{Total}$ & $12.6\pm0.8$ & $25.88\pm0.25$ & $23.53\pm0.16$ & $19.42\pm0.14$ \\ \hline

	{Disk} & $h$(") & $75\pm9$ & $50\pm5$ & $55\pm4$ & $45\pm3$  \\ 
	& $\mu_{0}$(mag/$arcsec^{2}$) & $22.3\pm0.1$ & $21.3\pm0.1$ & $21.3\pm0.1$ &$20.40\pm0.14$ \\ 

	& Ellipticity  & $0.67\pm0.05$ & $0.65\pm0.08$ & $0.68\pm0.05$ & $0.67\pm0.11$   \\ 
	& Position Angle  & $62\pm5$ & $63\pm5$ & $64\pm4$&$65\pm6$   \\
	& \% $L_{Disk}/L_{Total}$ & $83\pm2$ & $65\pm1$ & $66.5\pm0.7$ & $66.6\pm0.2$ \\

	& B/D & $0.15\pm0.03$ & $0.29\pm0.05$  & $0.34\pm0.024$ & $0.3\pm0.1$  \\  \hline
	{Ring } & $I_{or}$(mag/$arcsec^{2}$) & $21.3\pm0.1$ & $21.3\pm0.1$ & $20.9\pm0.1$ & $20.9\pm0.1$ \\ 
	& $r_{or}$(") & $21.6\pm0.5$&$21.1\pm0.5$&$20.5\pm0.5$&$20.5\pm0.5$ \\ 
	& $\sigma_{r}$(") & $8.1\pm0.5$&$6.5\pm0.5$&$6.2\pm0.5$&$5.4\pm0.5$\\ 
	& Ellipticity  &$0.48\pm0.09$  & $0.44\pm0.11$ & $0.56\pm0.14$ & $0.5\pm0.1$  \\ 
	& Position Angle  & $78\pm7$ & $80\pm8$ &$76\pm12$ & $82\pm9$   \\
	& \% $L_{Ring}/L_{Total}$ & $0.83\pm0.13$ & $0.52\pm0.05$ & $0.72\pm0.04$ & $0.33\pm0.02$ \\ \hline

	{Lens} & $\mu_{l0}$(mag/$arcsec^{2}$) & $23.4\pm0.1$ & $22.1\pm0.1$ & $21.9\pm0.1$& $20.7\pm0.1$ \\
	& $r_{ol}$ & $54\pm1$ & $51.3\pm0.5$ & $54\pm1$&$48.6\pm0.5$ \\ 
	& Position Angle & $71\pm3$&$71\pm3$  & $71\pm6$ & $74\pm2$ \\ 
	& \% $L_{Lens}/L_{Total}$ & $3.97\pm0.02$ & $8.23\pm0.06$ & $9.22\pm0.01$ & $14.67\pm0.01$ \\
\hline
	\end{tabular}
      \]
   \end{table*}

\begin{table*}
      \begin{center}
{\bf{Table 7.} Structural parameters of  NGC 7217.}
	\end{center}
      \[
	 \begin {tabular}{cccccc}
            \hline
	{Structure} & {Parameters} & { B} & { V}& { R} & { I}  \\ \hline 
	{Bulge} &   $r_{e}$(") & $11.21\pm0.19$ & $10.72\pm0.22$ & $6.96\pm0.08$ &$6.75\pm0.10$  \\   
        & $\mu_{e}$ (mag/$arcsec^{2}$) & $20.70\pm0.04$ & $20.03\pm0.02$ & $18.82\pm0.03$ &$18.21\pm0.02$ \\ 
	& n & 2.5 & 2.8  & 2.5&2.5   \\ 
	& \% $L_{Bulge}/L_{Total}$ & $21.30\pm0.23$ & $22.78\pm0.58$ & $15.11\pm0.05$ & $15.57\pm0.21$ \\ \hline

	{Disk} & $h$(") & $26.9\pm0.7$ & $26.68\pm0.17$ & $24.26\pm0.21$ & $22.9\pm0.7$   \\ 
	& $\mu_{0}$(mag/$arcsec^{2}$) & $20.04\pm0.08$ & $19.48\pm0.03$ & $18.51\pm0.03$ &$17.88\pm0.06$ \\ 

	& Ellipticity  & $0.09\pm0.20$ &$0.12\pm0.05$  & $0.12\pm0.09$ & $0.09\pm0.23$  \\ 
	& Position Angle  & $91\pm8$ & $92\pm12$ & $92\pm29$&$92\pm57$    \\  
	& \% $L_{Disk}/L_{Total}$ & $78.1\pm0.9$ & $76.6\pm0.9$ & $84.58\pm0.17$ & $84.11\pm0.48$ \\

	& B/D & $0.26\pm0.01$ & $0.29\pm0.01$  & $0.18\pm0.01$ & $0.18\pm0.01$ \\ \hline
	{Blue nuclear ring} & $I_{or}$(mag/$arcsec^{2}$) & $22.08\pm0.05$ & $21.13\pm0.02$ &  &   \\ 
	& $r_{or}$(") & $10.8\pm0.5$&$10.8\pm0.5$& & \\ 
	& $\sigma_{r}$(") & $2.7\pm0.5$&$2.7\pm0.5$& & \\ 
	& Ellipticity  & $0.91\pm0.03$ & $0.91\pm0.02$ &  &      \\ 
	& Position Angle  & $73\pm11$ & $78\pm9$ & &     \\ 
	& \% $L_{Nuclear \ ring}/L_{Total}$ & $0.07\pm0.01$ & $0.1\pm0.06$ & &  \\ \hline
	{Red nuclear ring} & $I_{or}$(mag/$arcsec^{2}$) &  &  & $20.33\pm0.05$ & $19.44\pm0.05$  \\ 
	& $r_{or}$(") & & &$8.1\pm0.5$&$8.1\pm0.5$ \\ 
	& $\sigma_{r}$(") & & &$2\pm1$&$2.7\pm0.5$ \\ 
	& Ellipticity  &  &  & $0.97\pm0.03$ & $0.92\pm0.02$     \\ 
	& Position Angle  &  &  &$75\pm25$ &$74\pm7$     \\ 
	& \% $L_{Nuclear \ ring}/L_{Total}$ &  &  & $0.06\pm0.01$& $0.12\pm0.02$ \\ \hline

	{Inner ring} & $I_{or}$(mag/$arcsec^{2}$) & $23.1\pm0.1$ & $22.35\pm0.07$ & &  \\ 
	& $r_{or}$(") & $31.3\pm0.5$&$30.2\pm0.5$& &  \\ 
	& $\sigma_{r}$(") & $2.7\pm0.5$&$5.4\pm0.5$ & &  \\ 
	& Ellipticity  & $0.83\pm0.03$ & $0.86\pm0.03$ &  &      \\ 
	& Position Angle  & $83\pm7$ & $83\pm8$ & &     \\
	& \% $L_{Inner \ ring}/L_{Total}$ & $0.08\pm0.02$ & $0.18\pm0.08$ & & \\ \hline
  
	{Outer ring} & $I_{or}$(mag/$arcsec^{2}$) & $23.4\pm0.2$ & $23.6\pm0.2$ & $23.0\pm0.1$&$22.9\pm0.1$  \\ 
	& $r_{or}$(") & $75.6\pm0.5$&$75.6\pm0.5$&$75.1\pm0.5$ &$75.6\pm0.5$  \\ 
	& $\sigma_{r}$(") & $8.1\pm0.5$&$10.8\pm0.5$ &$9.72\pm0.5$ &$10.8\pm0.5$  \\  
	& Ellipticity  & $0.85\pm0.09$ &$0.90\pm0.05$  & $0.89\pm0.13$ & $0.94\pm0.06$     \\ 
	& Position Angle  &$73\pm20$  & $83\pm15$ &$89\pm21$ &$69\pm27$     \\   
	& \% $L_{Outer \ ring}/L_{Total}$ & $0.44\pm0.03$ & $0.28\pm0.05$ & $0.24\pm0.06$ & $0.18\pm0.09$ \\
\hline
	\end{tabular}
      \]
   \end{table*}

\begin{table*}
      \begin{center}
	{\bf{Table 8.} Structural parameters of NGC 7479.}
	\end{center}
      \[
	 \begin {tabular}{ccccc}
            \hline
	{Structure} & {Parameters}  & { V}& { R} & { I}  \\ \hline 
	{Bulge} &   $r_{e}$(") & $7.22\pm0.09$ & $5.7\pm0.4$ & $5.3\pm0.4$ \\    
        & $\mu_{e}$ (mag/$arcsec^{2}$) & $20.45\pm0.19$ & $19.39\pm0.04$ & $18.77\pm0.09$ \\ 
	& n & 1 & 1  & 1   \\ 
	& \% $L_{Bulge}/L_{Total}$ & $11.32\pm0.18$ & $10.98\pm0.32$ & $9.68\pm0.08$  \\ \hline

	{Disk} & $h$(") & $45\pm5$ & $36\pm6$ & $40\pm3$   \\ 
	& $\mu_{0}$(mag/$arcsec^{2}$) & $21.86\pm0.19$ & $20.9\pm0.8$ & $20.44\pm0.14$  \\ 

	& Ellipticity  & $0.24\pm0.06$ & $0.26\pm0.05$ &$0.25\pm0.08$   \\ 
	& Position Angle  & $118\pm9$& $116\pm7$ &$119\pm11$   \\
	& \% $L_{Disk}/L_{Total}$ & $62.65\pm0.15$ & $57.05\pm0.15$ & $61.93\pm0.15$  \\
  
	& B/D & $0.17\pm0.02$ & $0.18\pm0.05$  & $0.15\pm0.04$   \\  \hline
	{Bar} & Major axis (Type) & Flat & Flat & Flat \\ 
	& $\mu_{0}$(mag/$arcsec^{2}$) & $20.83\pm0.01$ & $20.03\pm0.01$ & $19.39\pm0.01$ \\ 
	& $\alpha (")$ & $51.0\pm0.5$ & $51.0\pm0.5$ & $52.0\pm0.5$  \\ 
	& $\beta (")$ & $9.2\pm0.5$ & $9.2\pm0.5$ & $8.6\pm0.5$  \\ 
 	& Minor axis (Type) & Flat & Flat & Flat  \\ 
	& $\alpha (")$ & $19.0\pm1.0$ & $19.0\pm1.0$ & $17.5\pm0.5$  \\ 
	& $\beta (")$ & $9.8\pm0.5$ & $9.8\pm0.5$ & $9.2\pm0.5$  \\ 
	& Position Angle &$98\pm2$ &$99\pm2$ & $99\pm2$\\ 
	& Ellipticity  &$0.76\pm0.02$ &$0.74\pm0.02$ & $0.72\pm0.02$\\
	& \% $L_{Bar}/L_{Total}$ & $26.03\pm0.13$ & $31.95\pm0.11$ & $28.38\pm0.18$  \\
 \hline 
	\end{tabular}
      \]
   \end{table*}

\begin{table*}
      \begin{center}
{\bf{Table 9.} Structural parameters of NGC 7606.}
	\end{center}
      \[
	 \begin {tabular}{cccccc}
            \hline
	{Structure} & {Parameters} & { B} & { V}& { R} & { I}  \\ \hline 
	{Bulge} &   $r_{e}$(") & $5.83\pm0.11$ & $4.99\pm0.05$ & $7.4\pm0.2$ &$5.86\pm0.06$ \\   
        & $\mu_{e}$ (mag/$arcsec^{2}$) & $21.84\pm0.02$ & $20.73\pm0.01$ & $20.74\pm0.03$ &$19.74\pm0.02$ \\ 
	& n & 2.1 & 2.0  & 2.5&2.3   \\ 
	& \% $L_{Bulge}/L_{Total}$ & $3\pm1$ & $4\pm1$ & $5.84\pm0.24$ & $6\pm1$ \\ \hline

	{Disk} & $h$(") & $51\pm6$ & $43\pm5$ & $43\pm4$ & $40\pm4$   \\ 
	& $\mu_{0}$(mag/$arcsec^{2}$) & $21.85\pm0.25$ & $20.9\pm0.3$ & $20.39\pm0.25$ &$19.71\pm0.25$ \\ 

	& Ellipticity  & $0.59\pm0.03$ & $0.59\pm0.02$ & $0.59\pm0.02$ & $0.59\pm0.02$   \\ 
	& Position Angle  & $53\pm2$ & $56\pm2$ &$56\pm2$ & $56\pm2$   \\  
	& \% $L_{Disk}/L_{Total}$ & $96.8\pm0.7$ & $96\pm1$ & $94\pm2$ & $94.7\pm0.7$ \\

	& B/D & $0.034\pm0.001$ & $0.041\pm0.003$  & $0.067\pm0.007$ & $0.058\pm0.002$     \\  \hline
	\end{tabular}
      \]
   \end{table*}

\begin{table*}
      \begin{center}
	{\bf{Table 10.} Structural parameters of NGC 7723.}
	\end{center}
      \[
	 \begin {tabular}{cccccc}
            \hline
	{Structure} & {Parameters}  & { B}& { V} & { R}& { I}  \\ \hline 
	{Bulge} &   $r_{e}$(") & $3\pm1$ & $3.55\pm0.14$ & $3.79\pm0.04$ & $3.91\pm0.02$\\   
        & $\mu_{e}$ (mag/$arcsec^{2}$) & $20.66\pm0.06$ & $19.63\pm0.08$ & $19.46\pm0.04$& $18.49\pm0.01$ \\ 
	& n & 1 & 1  & 1 & 1  \\ 
	& \% $L_{Bulge}/L_{Total}$ & $5\pm4$ & $9\pm1$ & $8\pm1$ & $9.25\pm0.05$ \\ \hline

	{Disk} & $h$(") & $23\pm7$ & $22\pm4$ & $21\pm3$ & $22\pm5$  \\ 
	& $\mu_{0}$(mag/$arcsec^{2}$) & $21\pm2$ & $21\pm1$ & $20\pm1$ & $19.12\pm0.01$ \\ 

	& Ellipticity  & $0.31\pm0.08$ &$0.31\pm0.06$  &$0.32\pm0.08$ &  $0.31\pm0.02$ \\ 
	& Position Angle  & $127\pm3$&$129\pm7$  & $128\pm9$& $126\pm2$  \\  
	& \% $L_{Disk}/L_{Total}$ & $91\pm14$ & $82\pm10$ & $86\pm8$ & $86.1\pm0.7$ \\

	& B/D & $0.06\pm0.35$ & $0.1\pm0.1$  & $0.09\pm0.07$ & $0.1\pm0.1$   \\ \hline 
	{Bar} & Major axis (Type) & Elliptical & Elliptical & Elliptical & Elliptical\\ 
	& $\mu_{0}$(mag/$arcsec^{2}$) & $22.41\pm0.02$ & $21.15\pm0.01$ & $21.02\pm0.01$ & $20.3\pm0.01$\\ 
	& $a (")$ & $22\pm1$ & $22\pm1$ & $21\pm1$ & $23\pm1$ \\ 
 	& Minor axis (type) & Elliptical & Elliptical & Elliptical & Elliptical \\ 
	& $b (")$ & $11\pm1$ & $13\pm1$ & $14\pm1$ & $11\pm1$\\ 
	& Position angle &$159\pm4$ &$157\pm5$ & $157\pm4$& $156\pm3$\\ 
	& Ellipticity  &$0.5\pm0.1$ &$0.59\pm0.07$ &$0.66\pm0.07$ & $0.47\pm0.06$\\ 
	& \% $L_{Bar}/L_{Total}$ & $3.81\pm0.32$ & $8.87\pm0.08$ & $6.79\pm0.07$ & $5.02\pm0.04$ \\
\hline 
	\end{tabular}
      \]
   \end{table*}

\begin{table*}
      \begin{center}
	{\bf{Table 11.} Structural parameters of NGC 7753.}
	\end{center}
      \[
	 \begin {tabular}{ccccc}
            \hline
	{Structure} & {Parameters}  & { V}& { R} & { I}   \\ \hline 
	{Bulge} &   $r_{e}$(") & $4\pm1$ & $3\pm1$ & $4\pm1$  \\  
        & $\mu_{e}$ (mag/$arcsec^{2}$) & $20.64\pm0.12$ & $19.54\pm0.23$ & $19.45\pm0.09$   \\ 
	& n & 2 & 2  & 2   \\ 
	& $\% L_{Bulge}/L_{Total}$ & $9\pm2$ & $7\pm1$ & $12\pm2$ \\ \hline
	{Disk} & $h$(") & $22\pm3$ & $21\pm4$ & $20\pm3$   \\ 
	& $\mu_{0}$(mag/$arcsec^{2}$) & $20.64\pm$0.02 & $20.18\pm0.01$ & $19.64\pm0.01$  \\ 

	& Ellipticity  & $0.2\pm0.1$ & $0.2\pm0.1$ & $0.2\pm0.1$ \\ 
	& Position angle  & $150\pm10$&$147\pm11$  & $150\pm10$   \\  
	& $\% L_{Disk}/L_{Total}$ & $90\pm2$&$91\pm1$ & $85\pm3$  \\ 

	& B/D  & $0.09\pm0.04$  & $0.08\pm0.05$ & $0.14\pm0.09$     \\ \hline
	{Bar} & Type & Elliptical & Elliptical & Elliptical  \\ 
	& $\mu_{0}$(mag/$arcsec^{2}$) &$22.29\pm0.01$ & $21.20\pm0.03$ & $20.41\pm0.05$  \\ 
	& $a (")$ & $8.6\pm0.5$ & $10.8\pm0.5$ & $12.4\pm0.5$  \\ 
	& Position angle &$170\pm3$ &$171\pm2$ &$175\pm3$     \\ 
	& Ellipticity  &$0.45\pm0.03$ &$0.45\pm0.02$ &$0.44\pm0.03$ \\ 
	& $\% L_{Bar}/L_{Total}$ &$0.88\pm0.08$ &$2.24\pm0.12$ & $4.31\pm0.23$   \\ \hline  
	\end{tabular}
      \]
   \end{table*}

\clearpage

\begin{figure*}[!htb]
    \begin{flushleft}
\epsfig{figure=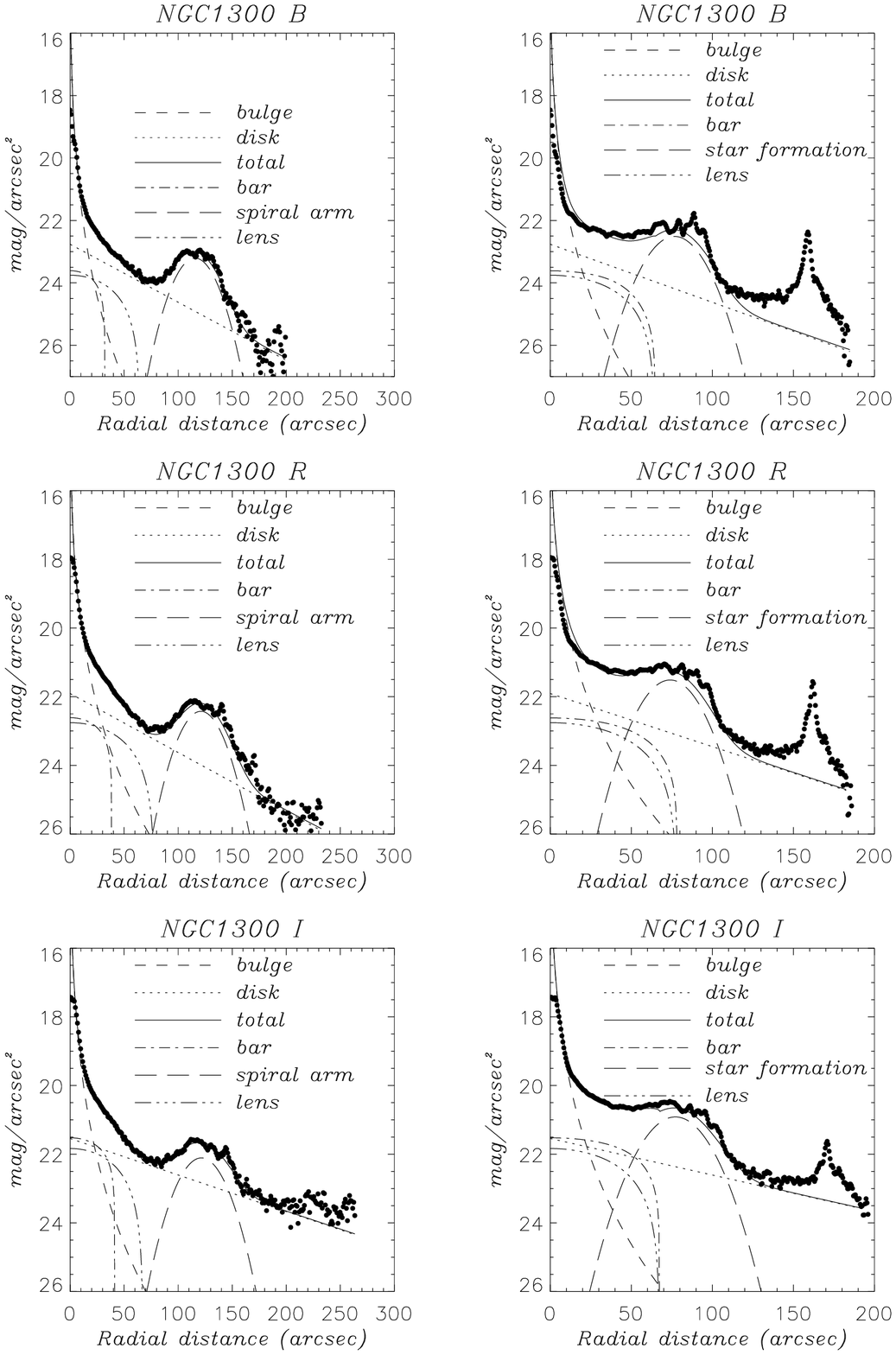,height=21.5truecm,width=17.truecm}
            {\bf{Fig 1.}}  {Structural decomposition of the surface-brightness profiles along the 
semi-minor bar axis (left) and semi-major bar axis (right) of NGC 1300 in $B$, $R$, 
and $I$.
}
    \end{flushleft}
    \end{figure*}

 \begin{figure*}[!htb]
    \begin{flushleft}
\epsfig{figure=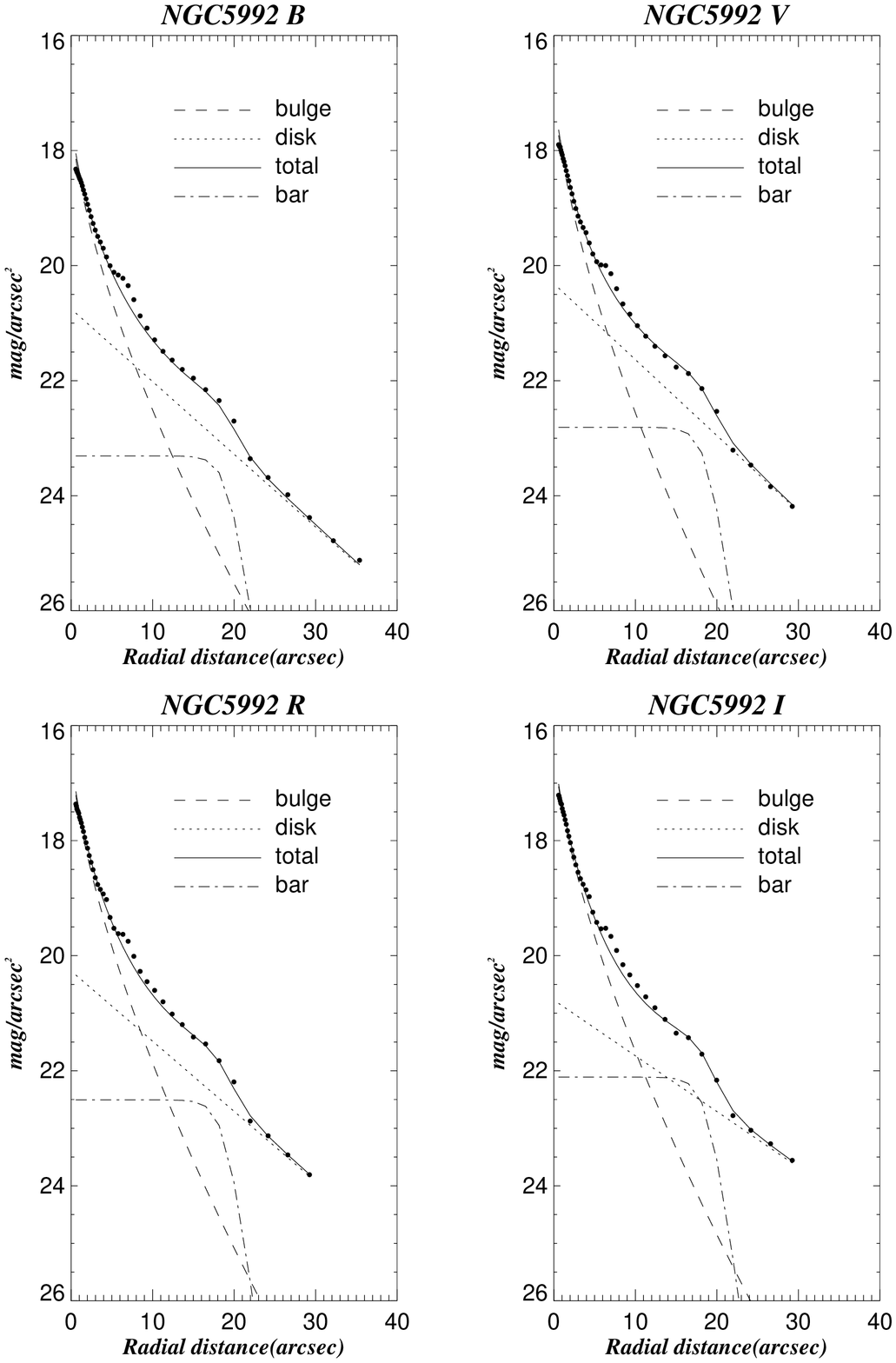,height=21.5truecm,width=17.truecm}
         {\bf{Fig 2.}}     { Structural decomposition of the average surface-brightness profiles of
 NGC 5992 in $B$, $V$, $R$, and $I$.}
    \end{flushleft}
    \end{figure*}

   \begin{figure*}[!htb]
    \begin{flushleft}
\epsfig{figure=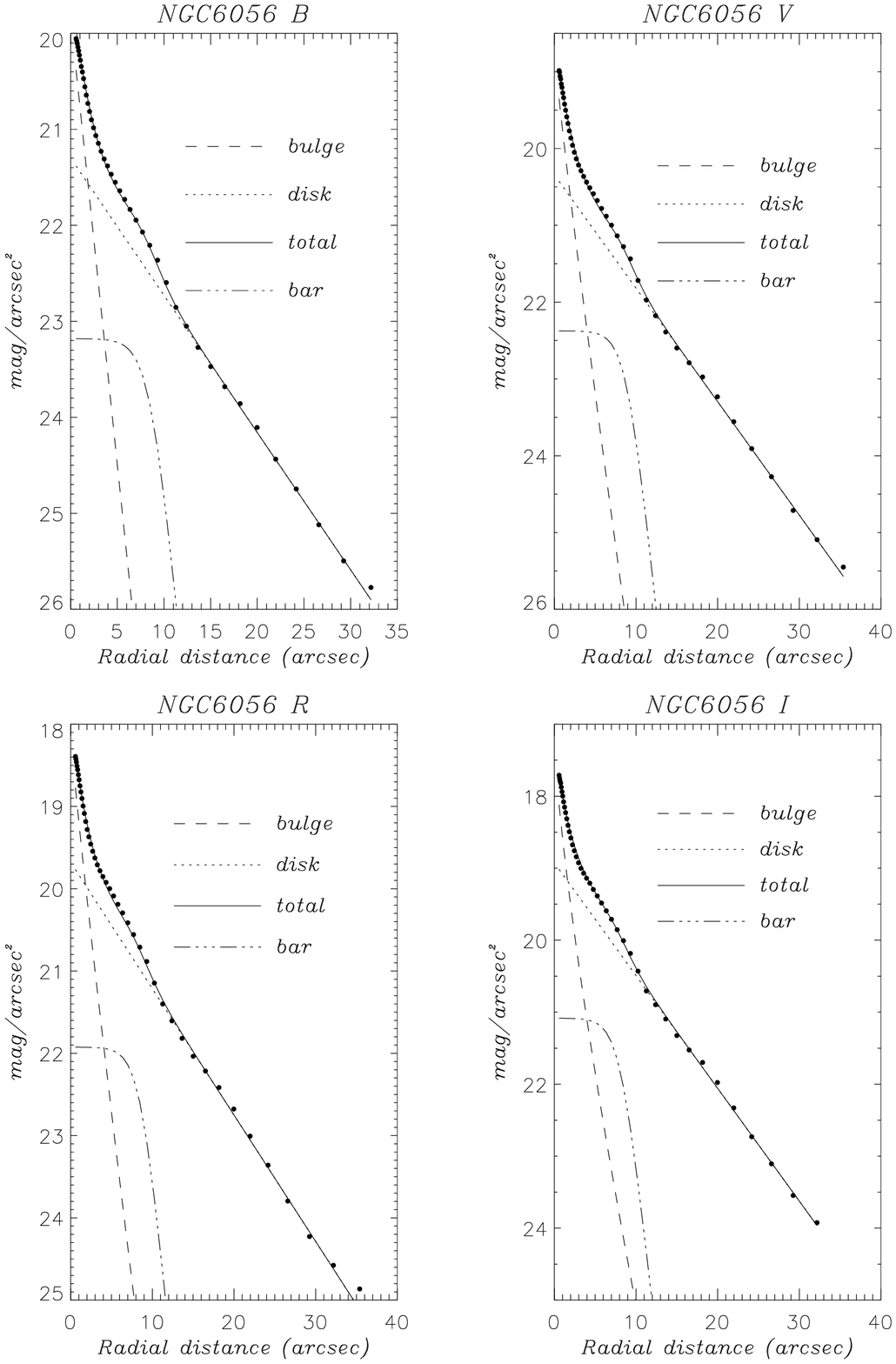,height=21.5truecm,width=17.truecm}
           {\bf{Fig 3.}}   {Structural decomposition of the average surface-brightness profiles of 
NGC 6056 in $B$, $V$, $R$, and $I$.}
    \end{flushleft}
    \end{figure*}

 \begin{figure*}[!htb]
    \begin{flushleft}
\epsfig{figure=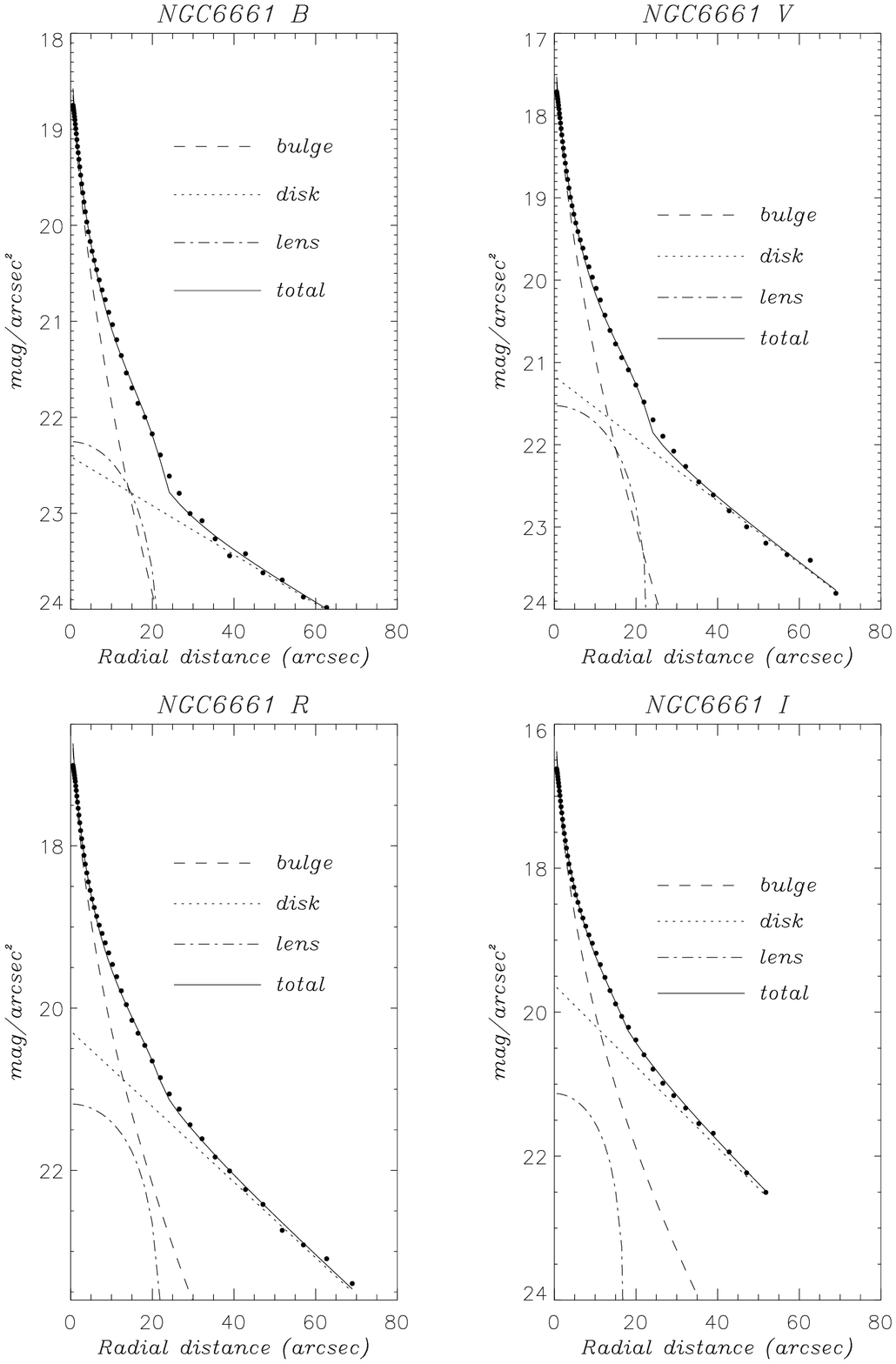,height=21.5truecm,width=17.truecm}
         {\bf{Fig 4.}}     {Structural decomposition of the average  surface-brightness profiles of 
NGC 6661 in $B$, $V$, $R$, and $I$.}
    \end{flushleft}
    \end{figure*}

 \begin{figure*}[!htb]
    \begin{flushleft}
\epsfig{figure=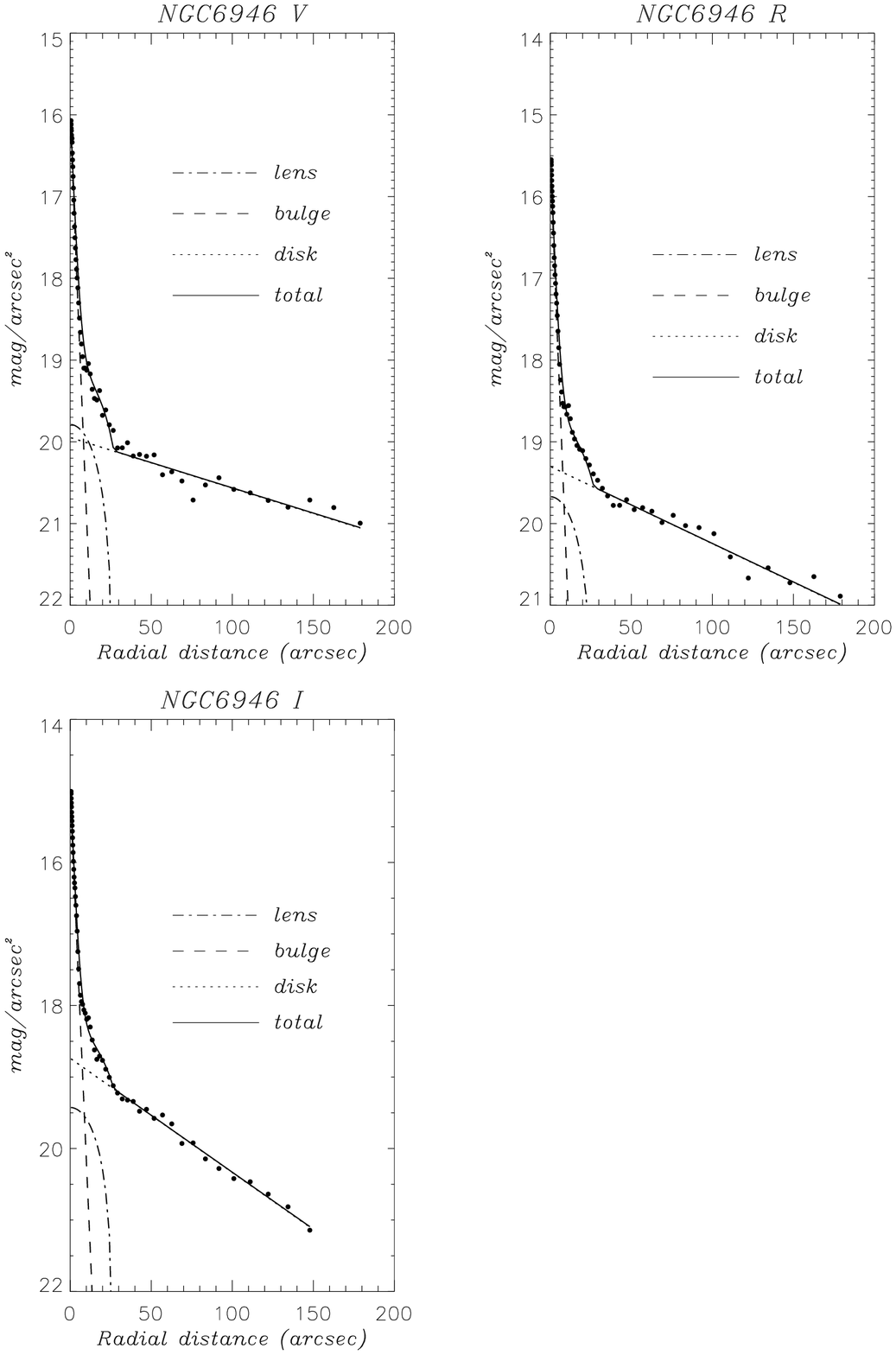,height=21.5truecm,width=17.truecm}
           {\bf{Fig 5.}}   {Structural decomposition of the average  surface-brightness profiles of 
NGC 6946 in $V$, $R$, and $I$.}
    \end{flushleft}
    \end{figure*}

 \begin{figure*}[!htb]
    \begin{flushleft}
\epsfig{figure=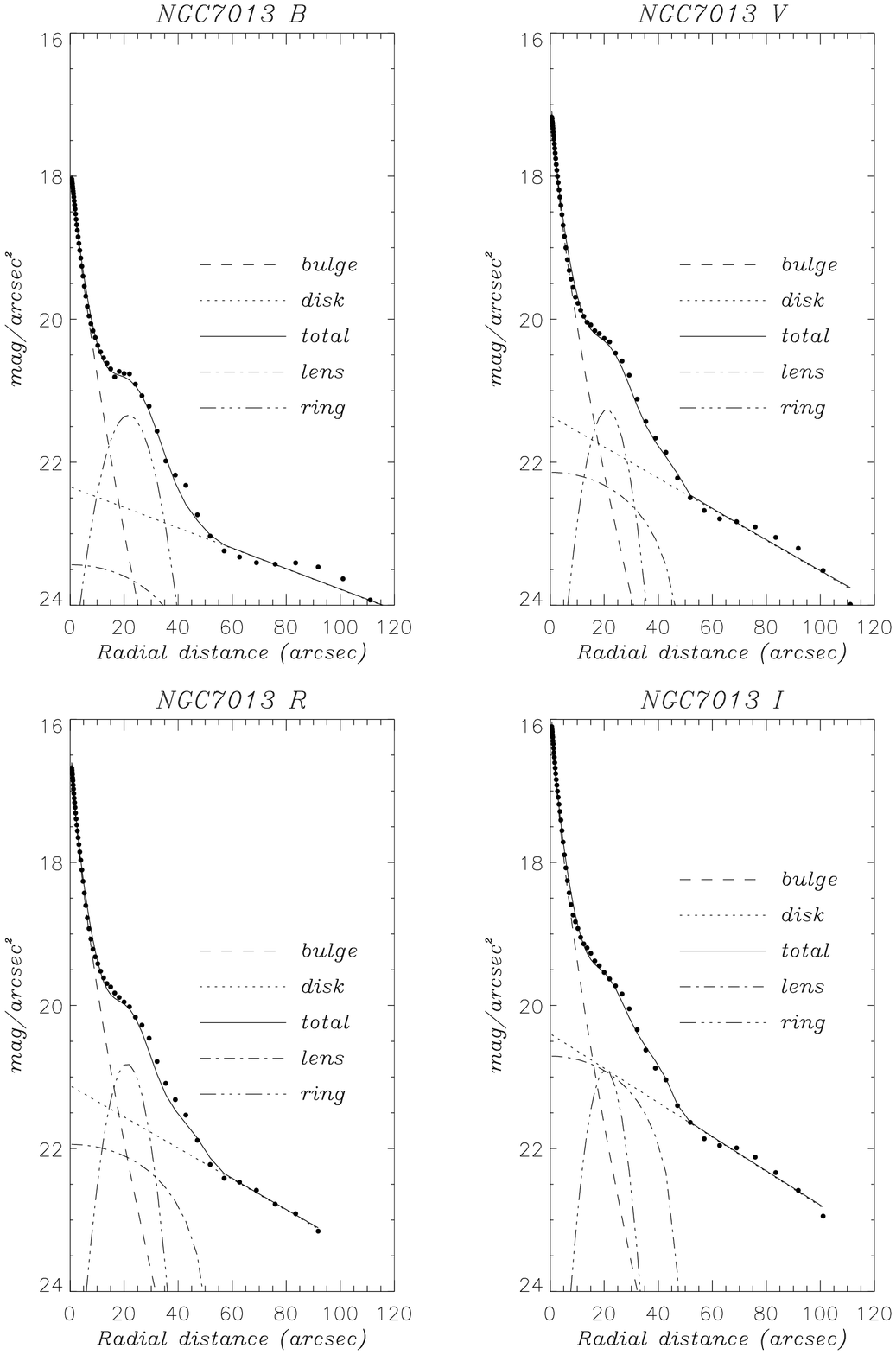,height=21.5truecm,width=17.truecm}
            {\bf{Fig 6.}}  { Structural decomposition of the  average surface-brightness profiles of 
NGC 7013 in $B$, $V$, $R$, and $I$.}
    \end{flushleft}
    \end{figure*}

 \begin{figure*}[!htb]
    \begin{flushleft}
\epsfig{figure=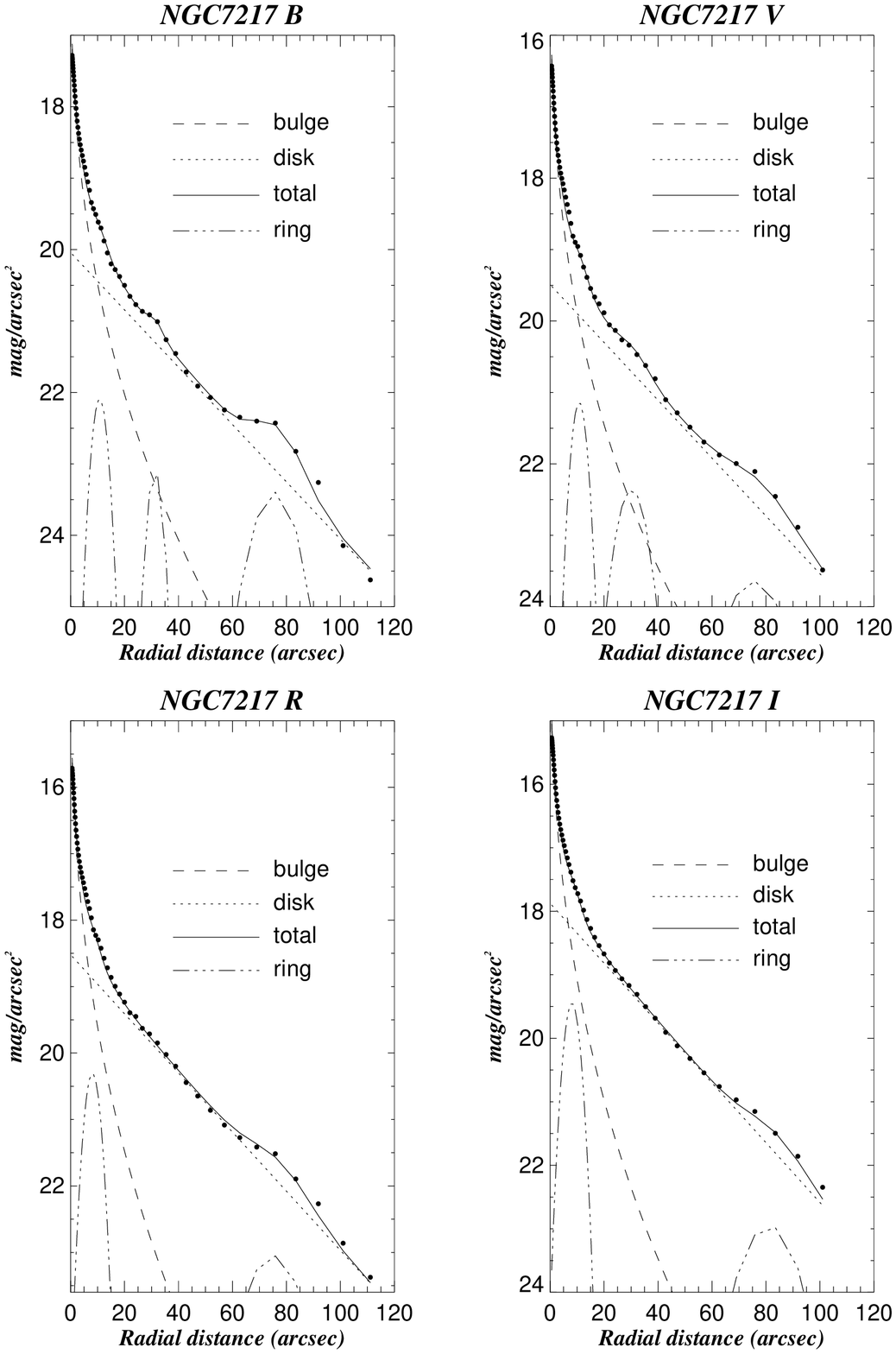,height=21.5truecm,width=17.truecm}
            {\bf{Fig 7.}}  { Structural decomposition of the average  surface-brightness profiles of 
NGC 7217 in $B$, $V$, $R$, and $I$.}
    \end{flushleft}
    \end{figure*}

  \begin{figure*}[!htb]
    \begin{flushleft}
\epsfig{figure=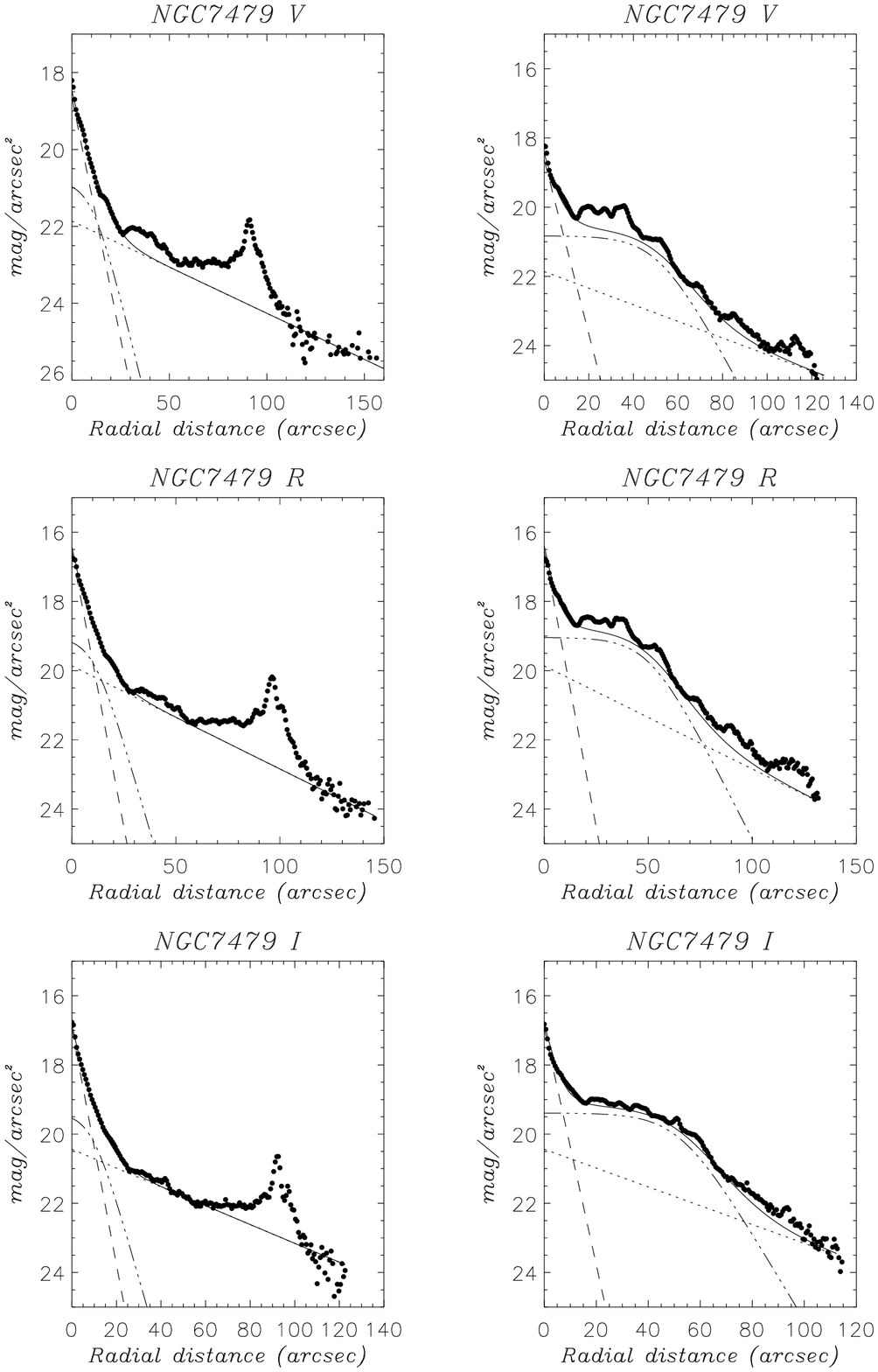,height=21.5truecm,width=17.truecm}
           {\bf{Fig 8.}}   {Structural decomposition of the surface-brightness profiles along the 
semi-minor bar axis (left) and semi-major bar axis (right) of NGC 7479 in $V$, $R$, 
and $I$.}
    \end{flushleft}
    \end{figure*}

 \begin{figure*}[!htb]
    \begin{flushleft}
\epsfig{figure=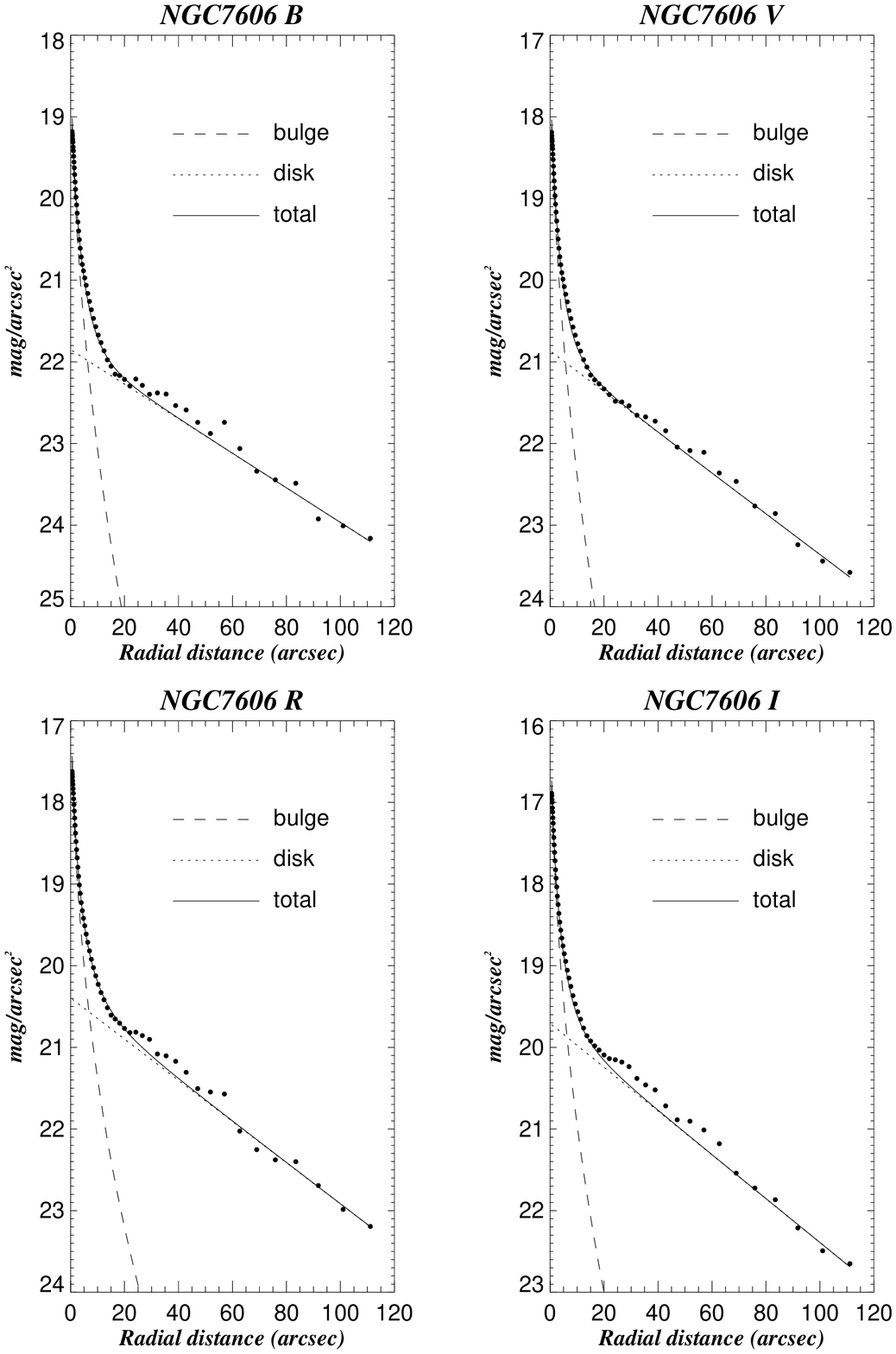,height=21.5truecm,width=17.truecm}
            {\bf{Fig 9.}}  { Structural decomposition of the average  surface-brightness profiles of
 NGC 7606 in $B$, $V$, $R$, and $I$.}
    \end{flushleft}
    \end{figure*}

  \begin{figure*}[!htb]
    \begin{flushleft}
\epsfig{figure=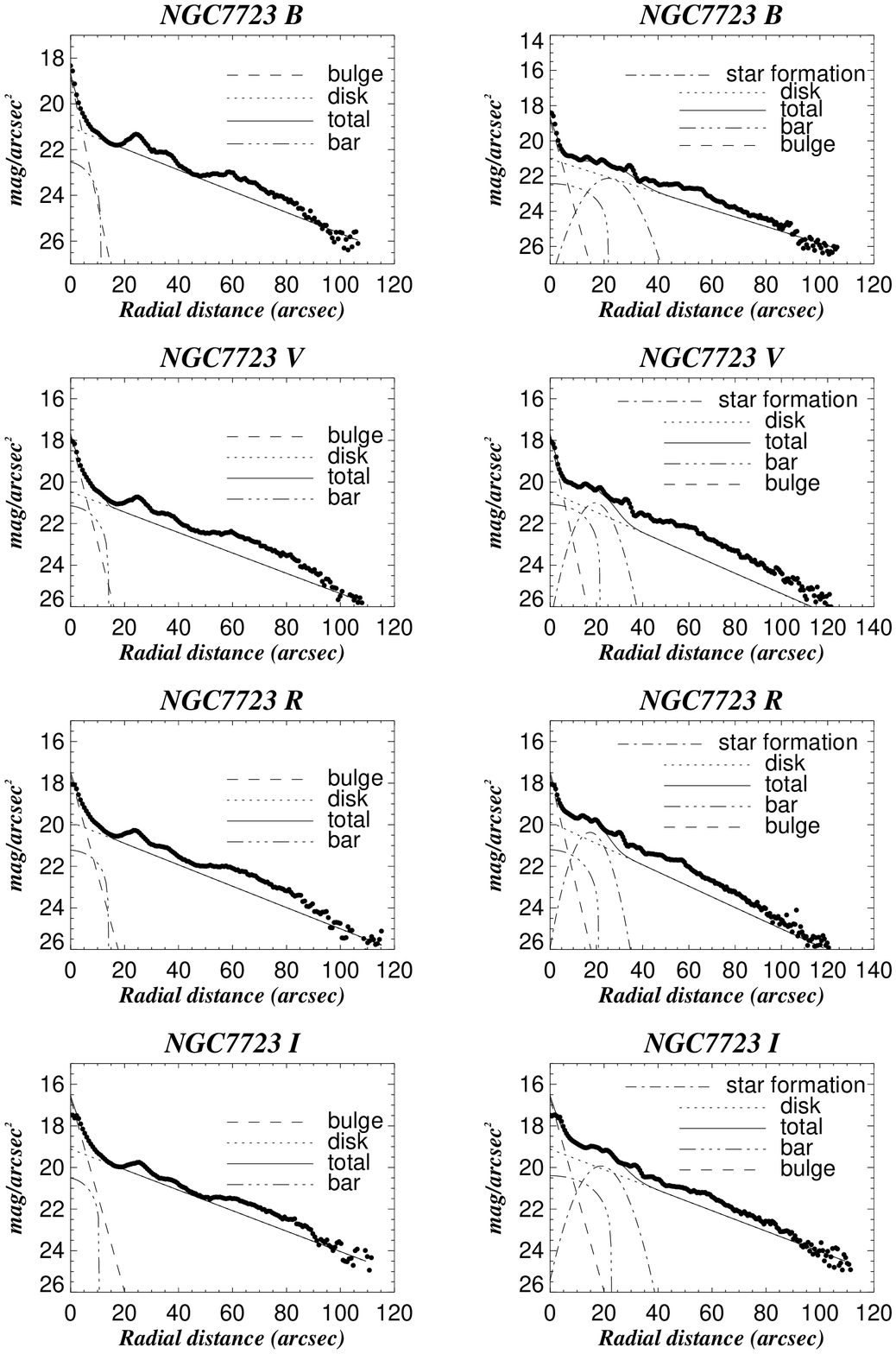,height=21.5truecm,width=17.truecm}
            {\bf{Fig 10.}}  { Structural decomposition of the surface-brightness profiles along the 
semi-minor bar axis (left) and semi-major bar axis (right) of NGC 7723 in $B$, $V$,
 $R$, and $I$.}
    \end{flushleft}
    \end{figure*}

 \begin{figure*}[!htb]
    \begin{flushleft}
\epsfig{figure=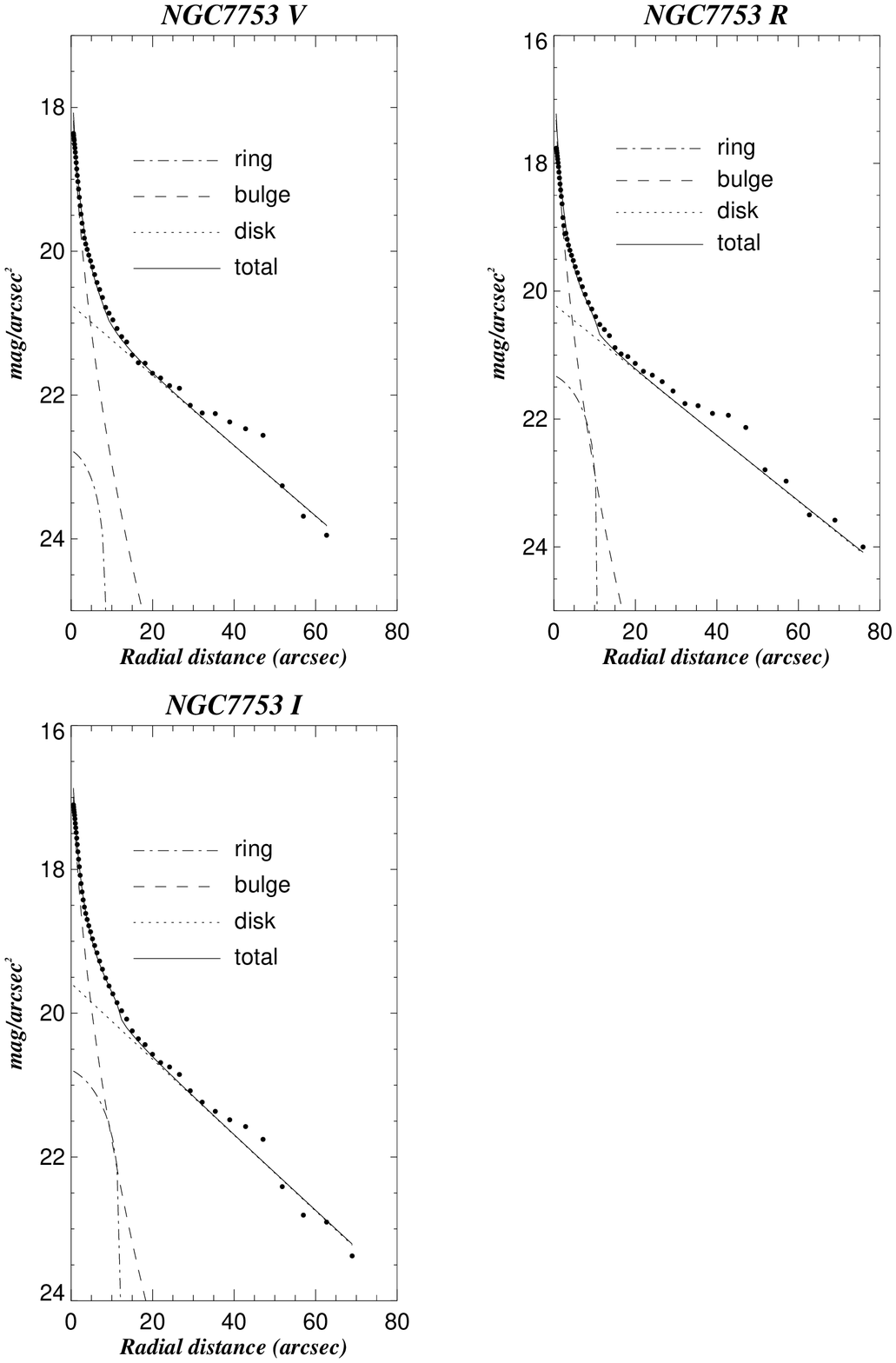,height=21.5truecm,width=17.truecm}
           {\bf{Fig 11.}}   { Structural decomposition of the average  surface-brightness profiles
 of NGC 7753 in $V$, $R$, and $I$.}
    \end{flushleft}
    \end{figure*}

\end{document}